\documentclass[aps,prb,reprint,superscriptaddress,longbibliography,floatfix]{revtex4-2}

\usepackage{graphicx}
\usepackage{amsmath,amssymb}
\usepackage{booktabs}
\usepackage[T1]{fontenc}
\usepackage[utf8]{inputenc}
\usepackage{xcolor}
\usepackage[colorlinks=true,linkcolor=blue,citecolor=blue,urlcolor=blue]{hyperref}

\begin{document}

\title{Multistable Thermoelectric Josephson Automata}

\author{Filippo Antola}
\email{filippo.antola@sns.it}
\affiliation{NEST, Istituto Nanoscienze-CNR and Scuola Normale Superiore, I-56127 Pisa, Italy}
\author{Margherita Cervi}
\affiliation{Dipartimento di Fisica, Universit\`a di Pisa, Largo Bruno Pontecorvo 3, I-56127 Pisa, Italy}
\author{Angelo Greco}
\affiliation{Istituto Nanoscienze-CNR, NEST-SNS, Piazza San Silvestro 12, I-56127 Pisa, Italy}
\author{Alessandro Braggio}
\affiliation{Istituto Nanoscienze-CNR, NEST-SNS, Piazza San Silvestro 12, I-56127 Pisa, Italy}
\author{Giorgio De Simoni}
\affiliation{Istituto Nanoscienze-CNR, NEST-SNS, Piazza San Silvestro 12, I-56127 Pisa, Italy}
\author{Francesco Giazotto}
\affiliation{Istituto Nanoscienze-CNR, NEST-SNS, Piazza San Silvestro 12, I-56127 Pisa, Italy}

\begin{abstract}
Multistable systems can store and process information, yet their elements usually only burn the drive energy. Here, we build an automaton whose elements actively convert heat into electrical power. Heat alone stabilizes the states at zero bias, and each state routes generation and dissipation to different junctions. The elements, gap-asymmetric superconducting tunnel junctions, deliver up to 2.6~pW at either polarity under fixed heating and support a magnetic-field-tunable Josephson current. Two junctions sharing a heated island split the bias hysteretically and, even at zero bias, lock into opposite voltages encoding the sweep direction. With a load and current bias, the automaton accesses states spanning junction polarity combinations, including zero voltage; the magnetic field sets which are available and the bias history selects the occupied one.
\end{abstract}

\maketitle

Coupled nonlinear elements may display coexisting stable configurations under an identical external drive, with the realized configuration determined by the system's prior trajectory (hysteresis)~\cite{BonillaGrahn2005,Kleiner1992,Heinrich2010}. For example, two elements in series with non-monotonic current--voltage relations can distribute a fixed applied voltage in more than one dynamically stable partition~\cite{BoricLubecke1996,Xu2010,Waldmann2026}. Such multistability provides the operating principle for Esaki-diode logic and multivalued memory concepts based on resonant tunneling~\cite{Goto1960,YanDeen1992}. In extended networks, the spatial voltage profile across the elements can encode the history of a global bias~\cite{Altman2026}. Consequently, the same input may induce different switching events depending on the pre-existing state, analogous to observations in arrays of coupled bistable mechanical elements~\cite{Keim2019,vanHecke2021,Kwakernaak2023,Liu2024,Paulsen2026}.

In the electrical circuits above, retaining the partitioned state requires continuous electrical driving: dissipative elements cannot sustain a nonzero voltage without current, so at zero applied bias, all partitions collapse to a single configuration. Thermoelectric elements partly relax this requirement, because a finite temperature difference can sustain a thermovoltage at vanishing current. At zero electrical bias and under fixed heating, a circuit of such elements may thus admit multiple stationary partitions, selected by the prior bias history. This mechanism can be fully leveraged only with elements whose thermoelectric response can assume either sign, i.e., intrinsically bipolar elements. In conventional thermoelectric converters, the polarity is typically fixed by the material properties and the orientation of the temperature gradient~\cite{Snyder2008,He2017}, and can be inverted only externally, for example, by electrostatic gating, magnetic switching, or optical excitation~\cite{Qin2021,Josefsson2018,Ozaeta2014,Kolenda2016,deAraujo2024,Fang2026}. Within a circuit, the polarities are thus prescribed rather than dynamically selected, and the thermovoltages combine constructively or destructively as dictated by the design~\cite{Montecucco2014,Bohnert2018}.

Superconducting tunnel junctions provide a versatile platform for measuring and controlling heat currents~\cite{Pekola2015,PekolaKarimi2021}, but electron--hole symmetry strongly suppresses their linear thermoelectric response~\cite{Bergeret2018,Arrachea2025}. When the two superconducting electrodes exhibit unequal energy gaps, however, intense heating of the larger-gap electrode can induce a nonlinear bipolar response~\cite{Marchegiani2020PRL,Germanese2022,Antola2026Review}: the junction may either dissipate or generate electrical power, with the polarity determined by the external circuit rather than by intrinsic material asymmetry. Under load, the junction may thus settle in one of two stable states of opposite polarity, storing a single bit of information~\cite{Marchegiani2020PRL,Germanese2022}. The same structure can also sustain a Cooper-pair supercurrent at zero voltage, providing a third dissipation-free operating state whose critical current is tunable by a magnetic field~\cite{Barone1982}. Here, we advance this device concept with an extended geometry and quasiparticle traps, which raise the output power from the femtowatt scale of earlier bipolar junctions and quantum-dot heat engines~\cite{Germanese2022,Josefsson2018,Jaliel2019} to the picowatt regime.

\begin{figure*}[t]
    \centering
    \includegraphics[width=\textwidth]{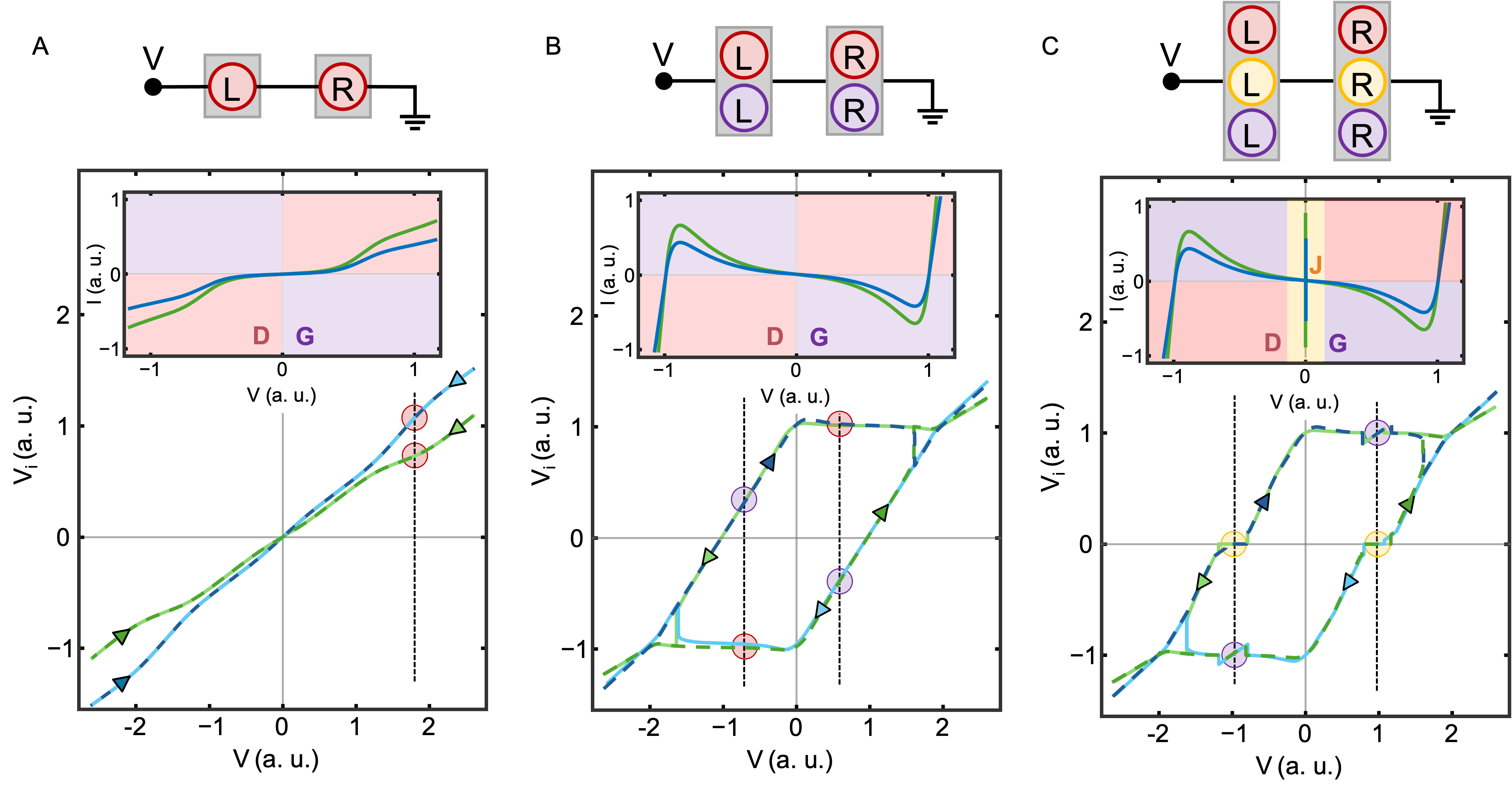}
\caption{\textbf{Operating principle: states and roles of two elements in series.}
Top: two elements, L and R, in series with a voltage source $V$. Circles in each box denote the roles accessible to that element: dissipation (D, red), generation (G, purple), and Josephson transport (J, yellow). Insets, current--voltage characteristics of L (blue) and R (green), with shading marking D ($IV>0$), G ($IV<0$), and J (zero-voltage branch). Main panels, local voltages $V_L$ and $V_R$ versus $V$, with light solid and dark dashed curves, and arrows, for sweeps from positive to negative bias and back. Circles at the bias points marked by vertical dashed lines indicate the role of each element. (A) Dissipative elements: a single partition at each bias. (B) Bipolar elements: the two sweeps form a hysteresis loop around $V=0$, inside which the local voltages have opposite signs, with equal magnitudes at $V=0$. (C) Bipolar elements with a zero-voltage branch: plateaus appear where one element sits at zero voltage (J). Curves are sketches in arbitrary units, and in (C) the zero-voltage plateaus are widened for visibility.}
\label{fig:principle}
\end{figure*}

Coupling two such junctions via a heated, electrically floating island introduces an internal node across which the applied bias is partitioned. The configuration of the two-junction system is then specified by the ordered pair of junction voltages, each positive, negative, or zero, and is selected by the prior bias trajectory. At zero applied bias, the system can reside in a configuration with equal-magnitude, opposite-polarity junction voltages and vanishing net current: the terminal voltage is zero, whereas the island potential records the direction of the preceding sweep. When current flows, the voltage configuration further determines the spatial distribution of energy conversion, i.e., which junctions convert heat into electrical power and which act predominantly as dissipative elements. We discriminate these contributions by directly probing the island potential. Under load and current bias, each stable configuration corresponds to an output level that heating can maintain even at zero bias, and the response to subsequent inputs depends on the driving history. The coupled pair therefore realizes a heat-powered \textit{automaton}, with accessible states and switching thresholds shaped by the magnetic field through the supercurrent.

\section*{Operating principle}
Two elements in series, driven by an external voltage $V$ (Fig.~\ref{fig:principle}, top), carry the same current, and their voltage drops add up to the bias $V=V_L+V_R$. Any pair $(V_L,V_R)$ satisfying both conditions is a possible partition.

Two slightly dissimilar dissipative elements with monotonic characteristics (inset) partition the bias in a single way, regardless of the sweep history (Fig.~\ref{fig:principle}A). Non-monotonic characteristics allow several stable partitions, or states of the pair, at the same bias, selected by the sweep direction~\cite{Heinrich2010,Altman2026}, but the drops of dissipative elements always share the sign of the current and vanish together at zero bias.

\begin{figure*}[t]
    \centering
    \includegraphics[width=\textwidth]{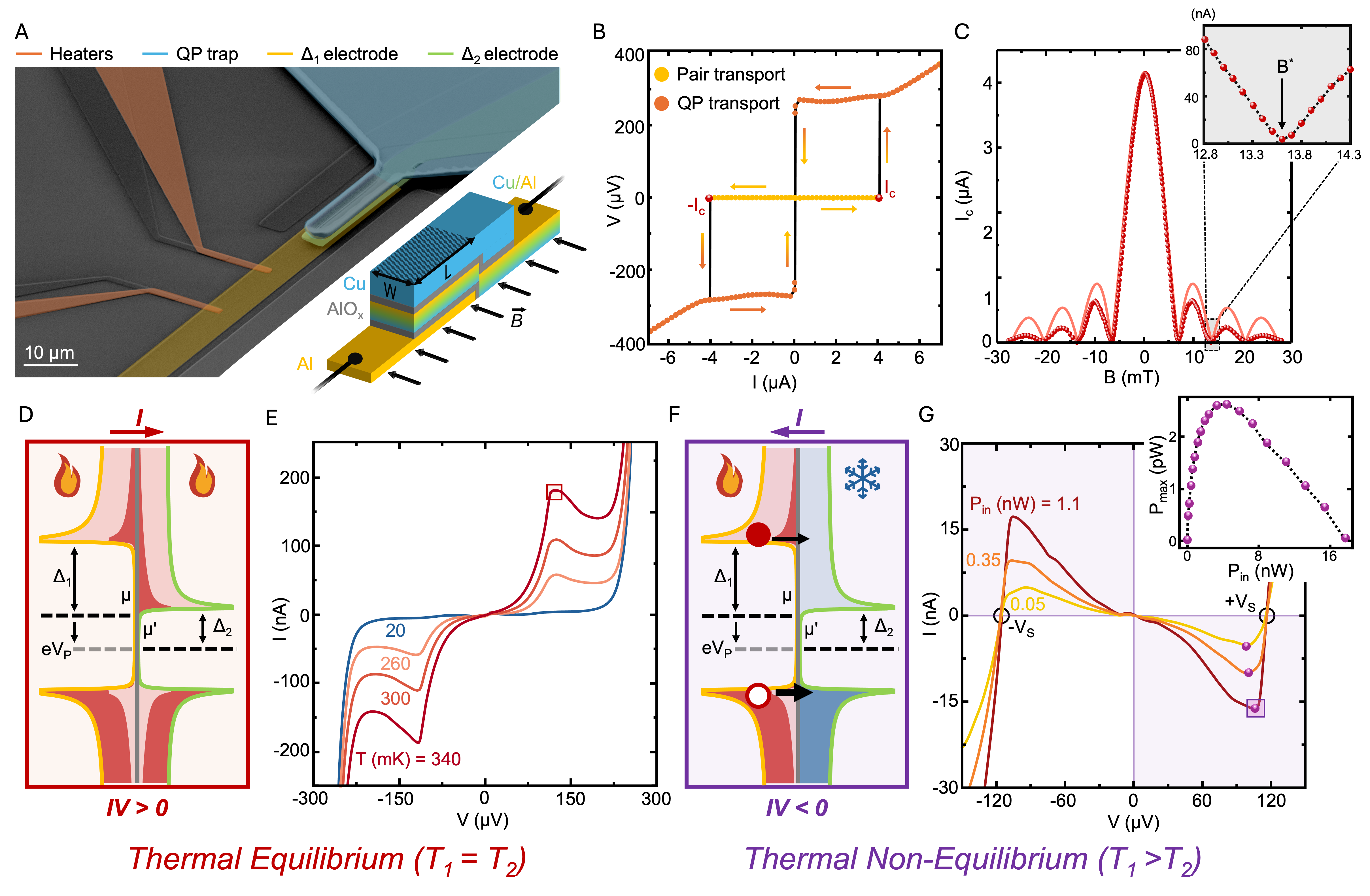}
\caption{\textbf{A picowatt bipolar thermoelectric junction.}
(A) False-color micrograph of the single-junction device (scale bar, 10~$\mu$m) and schematic of the junction stack, with black arrows indicating the in-plane magnetic field $B$.
(B) Current-biased $V(I)$ characteristic at zero magnetic field. Yellow and orange mark the pair and quasiparticle transport, and arrows indicate the sweep direction.
(C) Critical current $I_c$ versus $B$ (circles), with the ideal Fraunhofer pattern for the measured period (line). Inset, close-up of the second minimum, $B^*\simeq13.6$~mT.
(D and F) Schematic quasiparticle densities of states for $T_1=T_2$ (D) and $T_1>T_2$ (F). Shading marks occupied states and dashed lines mark the chemical potentials. In (F), filled and open circles denote electron-like and hole-like excitations, and black arrows denote the associated tunneling processes. Arrows above the schematics indicate the net charge current.
(E and G) Current--voltage characteristics at $B^*$ for the indicated equilibrium temperatures (E) and heating powers $P_{\rm in}$ (G). Squares mark the matching conditions of (D) and (F). In (G), purple shading highlights generation ($IV<0$), purple dots the generation peaks, where $P_{\rm max}=-V_PI(V_P)$ is evaluated, and open circles the thermovoltages $\pm V_S$. Inset, $P_{\rm max}$ versus $P_{\rm in}$.}
        \label{fig:block}
\end{figure*}
Instead, bipolar thermoelectric elements under a sufficiently large temperature difference admit opposite drops (Fig.~\ref{fig:principle}B). Along a sweep, the less conductive element swings, taking up most of the change in bias, while the other holds a nearly constant voltage. Since each element sustains a thermovoltage at zero current, the swing continues through zero bias, where, for equal thermovoltages, the drops are finite and opposite, and no current flows. Past it, the other element takes up the rest of the sweep. As the less conductive element swings first in both directions, descending and ascending sweeps reach zero bias with opposite signs of $V_L$ and $V_R$. The hysteresis loop thus encloses zero bias, leaving the pair in one of two mirror-image states, labeled $(+,-)$ and $(-,+)$ by these signs. Indistinguishable at the terminals, they store the sweep memory in the potential of the node between them.

When current flows, the sign of $IV_i$ identifies the role of each element: dissipative (D) if positive, generative (G) if negative. Concordant drops make the two elements generate or dissipate together, whereas opposite drops make one generate and the other dissipate. By selecting the state, the history thus selects the roles: within the loop, at the same bias, the two sweeps give D-G and G-D (Fig.~\ref{fig:principle}B). An element with a zero-voltage branch adds the dissipationless (J) role, exchanging no power and leaving the whole bias to its partner. The local voltages then show plateaus where one element sits at zero voltage, labeled 0, and which one again depends on the sweep direction (Fig.~\ref{fig:principle}C).

\section*{The bipolar thermoelectric element}
The operating principle of Fig.~\ref{fig:principle} requires elements that generate power at either polarity under the same heating and carry a zero-voltage branch of tunable extent. Figure~\ref{fig:block}A shows an extended Al/AlO$_x$/Cu/Al tunnel junction ($3.5\times15~\mu\mathrm{m}^2$, $R_N\simeq37~\Omega$), whose Al electrode is heated by two floating tunnel heaters with power $P_{\rm in}$. 
The insulating barrier faces the Cu side of the Cu/Al bilayer, where proximity induces a gap $\Delta_2$ smaller than the Al electrode gap $\Delta_1$. This S/I/NS configuration is predicted~\cite{hijano2023} to favor bipolar thermoelectricity over the S/I/SN geometry of earlier devices~\cite{Germanese2022}. The large area lowers $R_N$ and thus increases the generated power~\cite{Benenti2017}, but also carries more heat into the bilayer, where superconductivity suppresses electron--phonon relaxation. A tunnel-coupled Cu quasiparticle trap therefore drains this heat and keeps the bilayer cold~\cite{Giazotto2006}.

The junction carries supercurrent up to $I_c\simeq4.1~\mu$A (J role) and then switches to the dissipative branch (D role) near the sum-gap threshold $V_\Sigma=(\Delta_1+\Delta_2)/e$ (Fig.~\ref{fig:block}B). The long side of the junction makes $I_c$ sensitive to an in-plane magnetic field, which modulates it through a Fraunhofer-like interference pattern (Fig.~\ref{fig:block}C)~\cite{Rowell1963} and thus sets the extent of the J branch (Fig.~\ref{fig:principle}C). Near $B^*\simeq13.6$~mT, $I_c$ falls by nearly three orders of magnitude (inset) without appreciably modifying the gaps~\cite{methods}.

\begin{figure*}[t]
    \centering
    \includegraphics[width=\textwidth]{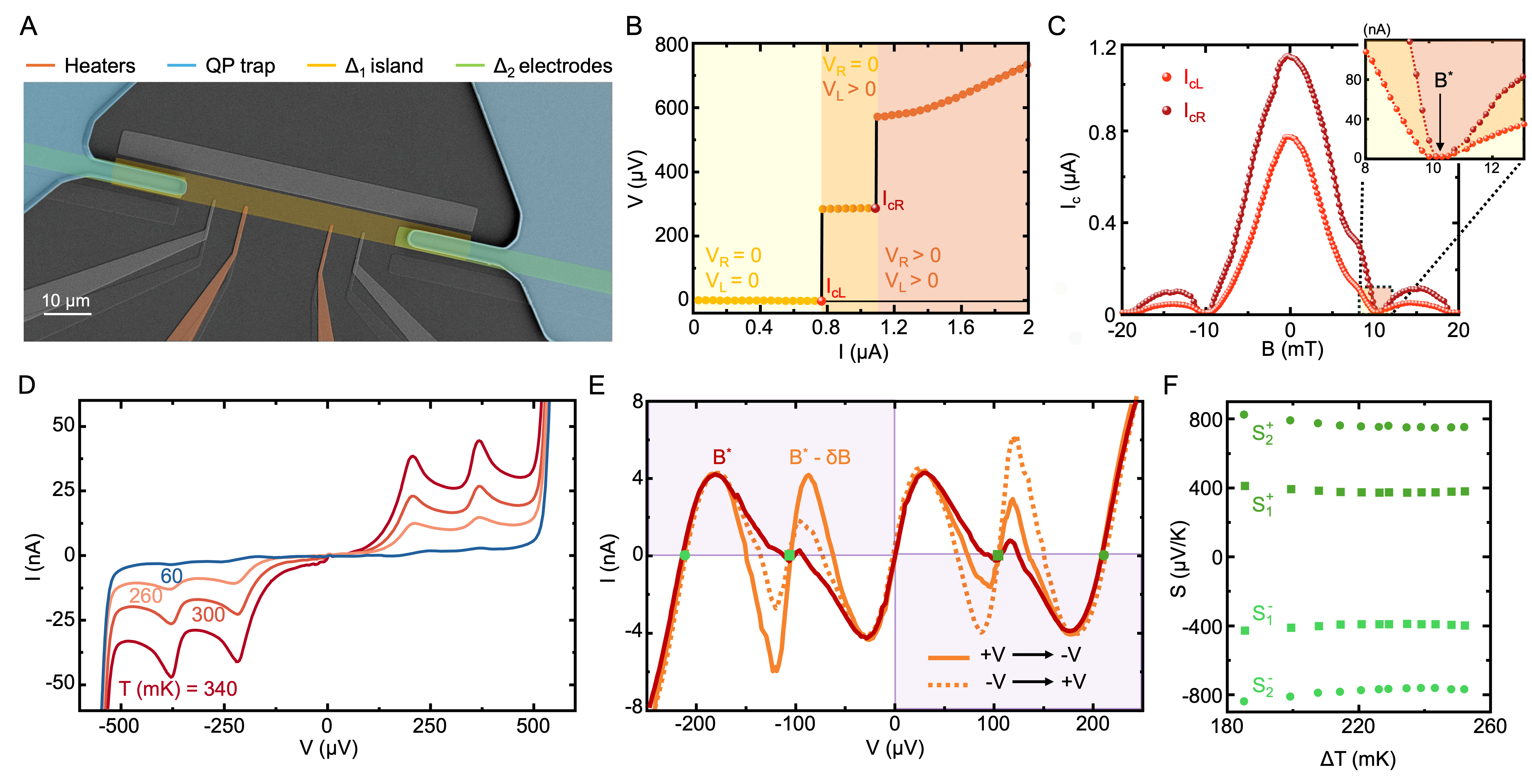}
\caption{\textbf{Two junctions sharing a hot island.}
(A) False-color micrograph of the two-junction device, with a shared floating Al island coupled to two Cu/Al electrodes (scale bar, 10~$\mu$m). A second pair of heaters is left uncolored, and the gray bar above the island is a shadow-mask residue.
(B) Current-biased $V(I)$ characteristic at zero magnetic field. Shaded regions delineate three regimes: both junctions in the zero-voltage state, only L switched (the junction that switches first), and both switched.
(C) Critical currents $I_{cL}$ and $I_{cR}$ versus $B$. Inset, common minimum near $B^*\simeq10.1$~mT, with shading as in (B).
(D) Current--voltage characteristics at equilibrium at $B^*$ for the indicated temperatures.
(E) Current--voltage characteristics under heating at $P_{\rm in}=0.7$~nW, at $B^*$ (red) and at $B^*-\delta B$, with $\delta B=1.4$~mT (orange). Purple shading highlights generation ($IV<0$). Solid and dotted lines denote sweeps from positive to negative bias and back. At $B^*$, only the former is shown. Green squares and circles mark the inner ($\pm V_S$) and outer ($\pm 2V_S$) zero-current crossings.
(F) Inner ($S^\pm_1$, squares) and outer ($S^\pm_2$, circles) nonlinear thermopowers versus $\Delta T$, at $B^*$ and low heating ($P_{\rm in}\le4$~pW).}
    \label{fig:pair}
\end{figure*}

Suppressing the Josephson current at $B^*$ gives full access to the subgap response. Its shape is determined by the BCS density-of-states singularities at $\pm\Delta_1$ and $\pm\Delta_2$, which act as energy filters for tunneling (Fig.~\ref{fig:block}D). A voltage shifts one spectrum relative to the other by $eV$, and aligning the singularities at $e|V_P|\simeq\Delta_1-\Delta_2$ produces a current maximum, the matching peak. Beyond it, the current decreases, giving a negative differential conductance~\cite{Nicol1960,Taylor1963}. In equilibrium ($T_1=T_2$), the matching peak grows with temperature as the thermally excited quasiparticles increase (Fig.~\ref{fig:block}E). The junction is thus non-monotonic but dissipative (D).

Heating the Al electrode, while the trap keeps the bilayer cold ($T_1>T_2$), makes the tunneling of electron- and hole-like quasiparticles unequal (Fig.~\ref{fig:block}F). The resulting current can flow against the applied voltage, an absolute negative conductance~\cite{Marchegiani2020PRL}: the junction generates power (G). Reversing the bias exchanges the roles of electron- and hole-like quasiparticles, so that the junction generates power at either polarity under the same heating.
Figure~\ref{fig:block}G shows this at $B^*$: $IV<0$ at both polarities, with zero-current crossings at opposite thermovoltages $\pm V_S$. At the generation peak, the junction delivers $P_{\rm max}\simeq2.6$~pW, a power that decreases with stronger heating as the bilayer warms (inset). Thus, a single junction provides all three roles of Fig.~\ref{fig:principle}: dissipation, picowatt generation at either polarity, and a field-tunable Josephson branch.

\begin{figure*}[t]
    \centering
    \includegraphics[width=0.9\textwidth]{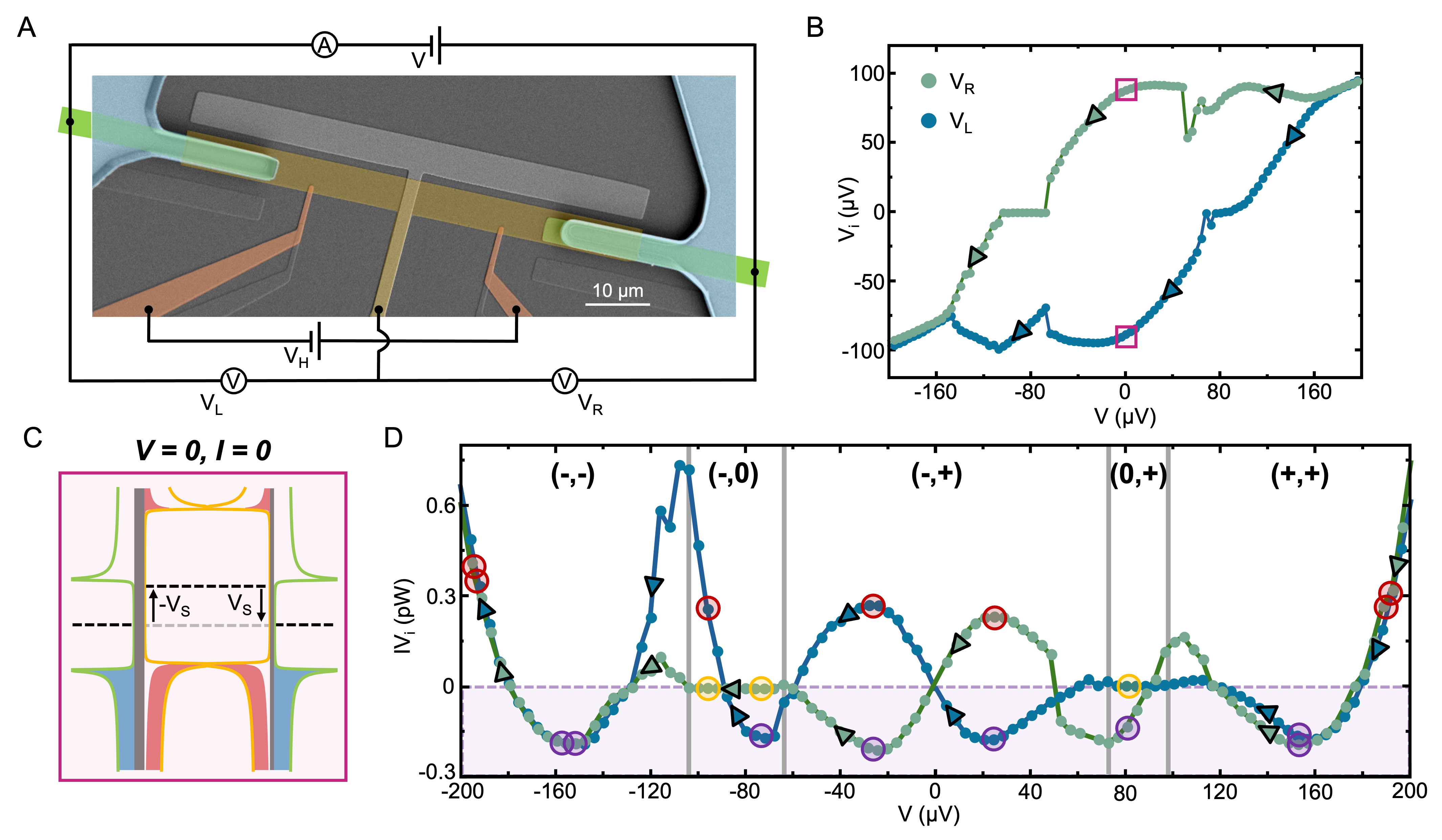}
    \caption{\textbf{States and roles resolved at the island.}
(A) False-color micrograph of the device and measurement circuit (scale bar, 10~$\mu$m). A superconducting lead gives access to the island, and $V_H$ drives the on-chip heaters. Uncolored structures are shadow-mask residues.
(B) Measured $V_L$ (blue) and $V_R$ (green) versus $V$ at $B^*\simeq9$~mT and $P_{\rm in}=1.9$~nW, during a sweep from positive to negative bias (arrows). Squares mark $V=0$, where the pair rests in $(-,+)$.
(C) Density-of-states schematic of this state at $V=0$ and $I=0$, with thermovoltages $\pm V_S$ across the two junctions.
(D) Local powers $P_i=IV_i$ for the same sweep, positive for dissipation and negative for generation, with purple shading. Vertical lines separate the five states, labeled by the signs of $V_L$ and $V_R$, and circles mark dissipation (red), generation (purple), and Josephson transport (yellow).}
    \label{fig:local}
\end{figure*}

\section*{Two junctions sharing a hot island}
Figure~\ref{fig:pair}A shows two nominally identical junctions, with a geometry similar to that of Fig.~\ref{fig:block}~\cite{methods}, connected in series through an electrically floating Al island that on-chip heaters warm as their common hot electrode. In zero field, the characteristic shows two successive switches out of the Josephson state, at $I_{cL}\simeq0.8~\mu$A and $I_{cR}\simeq1.1~\mu$A (Fig.~\ref{fig:pair}B). The lower critical current of L suggests a higher normal-state resistance within fabrication tolerances. Near their common interference minimum, $B^*\simeq10.1$~mT, both critical currents fall to the nanoampere scale (Fig.~\ref{fig:pair}C), exposing the quasiparticle response of the pair.

Already at equilibrium, the bias partition is non-trivial (Fig.~\ref{fig:pair}D): instead of a single subgap peak near $2V_P$, two peaks emerge as the bath temperature increases, one for each junction reaching its matching peak. 

Heating the island at $B^*$ makes the pair generate: its characteristic develops branches with $IV<0$ (Fig.~\ref{fig:pair}E) and, besides $V=0$, crosses zero current near $\pm V_S$ and $\pm2V_S$. At the inner crossings, one junction holds its thermovoltage $\pm V_S$ (Fig.~\ref{fig:block}G) while the other sits at zero voltage, and at the outer crossings, the two thermovoltages have the same sign and add. Observed at both polarities, together with generation peaks near $\pm2V_P$, this addition further confirms the bipolar nature of thermoelectric generation.

Slightly detuning the field from $B^*$ reopens the critical currents and makes the J role visible (Fig.~\ref{fig:pair}E, orange). Near the inner crossings, the descending sweep (solid) then shows a complete single-junction generation peak near $-V_P$,  while at positive bias the generated current reaches only about half that value. Since R's critical current reopens first (Fig.~\ref{fig:pair}C, inset), the complete peak at negative bias corresponds to L generating while R carries a supercurrent, and at positive bias the reverse. The ascending sweep (dotted) mirrors the characteristic, exchanging the polarities as in Fig.~\ref{fig:principle}C: the bias history selects which junction sits at zero voltage at each polarity.

Fitting the quasiparticle characteristics gives the island and electrode temperatures, and with them the thermoelectric performance of the pair (supplementary text S1). Dividing the crossing voltages by the temperature difference $\Delta T$ gives the nonlinear thermopowers~\cite{Marchegiani2020PRL} $S_n^\pm=V_n^\pm/\Delta T$, with $V_1^\pm\simeq\pm V_S$ and $V_2^\pm\simeq\pm2V_S$ the inner and outer zero-current crossings at both polarities (Fig.~\ref{fig:pair}F). The inner one, $|S_1^\pm|\simeq400~\mu$V\,K$^{-1}$, matches that of the single junction, and the outer one doubles it. The island is also designed for efficient conversion: small volume, superconducting, and contacted only through tunnel barriers, it loses heat mainly through the junctions rather than to lattice phonons. A few picowatts of heater power thus suffice: at low heating ($P_{\rm in}\simeq2$--$3$~pW), the pair converts about 3\% of it into electrical power, up to 90~fW. Since it refers to the supplied power, this figure is a lower bound on the thermodynamic efficiency, usually estimated in mesoscopic heat engines from calculated heat currents~\cite{Josefsson2018}.

\section*{Probing the island}
So far, the states have been inferred from the total voltage $V=V_L+V_R$ alone. Testing them requires electrical access to the island, which a second device provides at the cost of lower heating efficiency (Fig.~\ref{fig:local}A): a superconducting lead connects the island to high-impedance probes that measure $V_L$ and $V_R$ separately. The device's total characteristic is compatible with Fig.~\ref{fig:pair}E~\cite{methods}.

Along a descending sweep, the local voltages follow Fig.~\ref{fig:principle}C step by step (Fig.~\ref{fig:local}B). From high bias, where both voltages are positive, L takes up the change in bias, resting at zero on its Josephson branch before turning negative, while $V_R$ stays nearly constant. At $V=0$ no current flows, yet $V_R=-V_L\simeq V_S$: the pair rests in the mirror-image state $(-,+)$ of Fig.~\ref{fig:principle}B, and the island, offset by $V_S$ from both electrodes, retains the sweep memory (Fig.~\ref{fig:local}C). Past $V=0$, R takes over in the same way, through its own Josephson branch, until both voltages are negative. L leads because it is the less conductive junction, although identical junctions would also split the bias unequally. In the reverse sweep, the polarities exchange, as the mirrored characteristics of Fig.~\ref{fig:pair}E and the circuit model indicate (supplementary text S2).

With each junction labeled by its voltage sign, or 0 on its Josephson branch, the sweep becomes a sequence of five states, $(+,+)$, $(0,+)$, $(-,+)$, $(-,0)$ and $(-,-)$ (Fig.~\ref{fig:local}D). The local powers $P_i=IV_i$ show that each state distributes the conversion as anticipated in Fig.~\ref{fig:principle}: the junctions generate or dissipate together in the concordant states, one does so alone when its partner sits at zero, and they oppose each other in $(-,+)$. In each state, generation turns into dissipation where the current reverses, at the zero-current crossings of Fig.~\ref{fig:pair}E, so that, with the Josephson plateaus, each junction takes all three roles within a single sweep. The reverse sweep visits the mirror states, $(+,0)$, $(+,-)$ and $(0,-)$, which differ from their counterparts not in kind but in memory: the island potential, and the role each junction plays at a given bias, record the direction of the last sweep.
\begin{figure}[!tb]
    \centering
    \includegraphics[width=\linewidth]{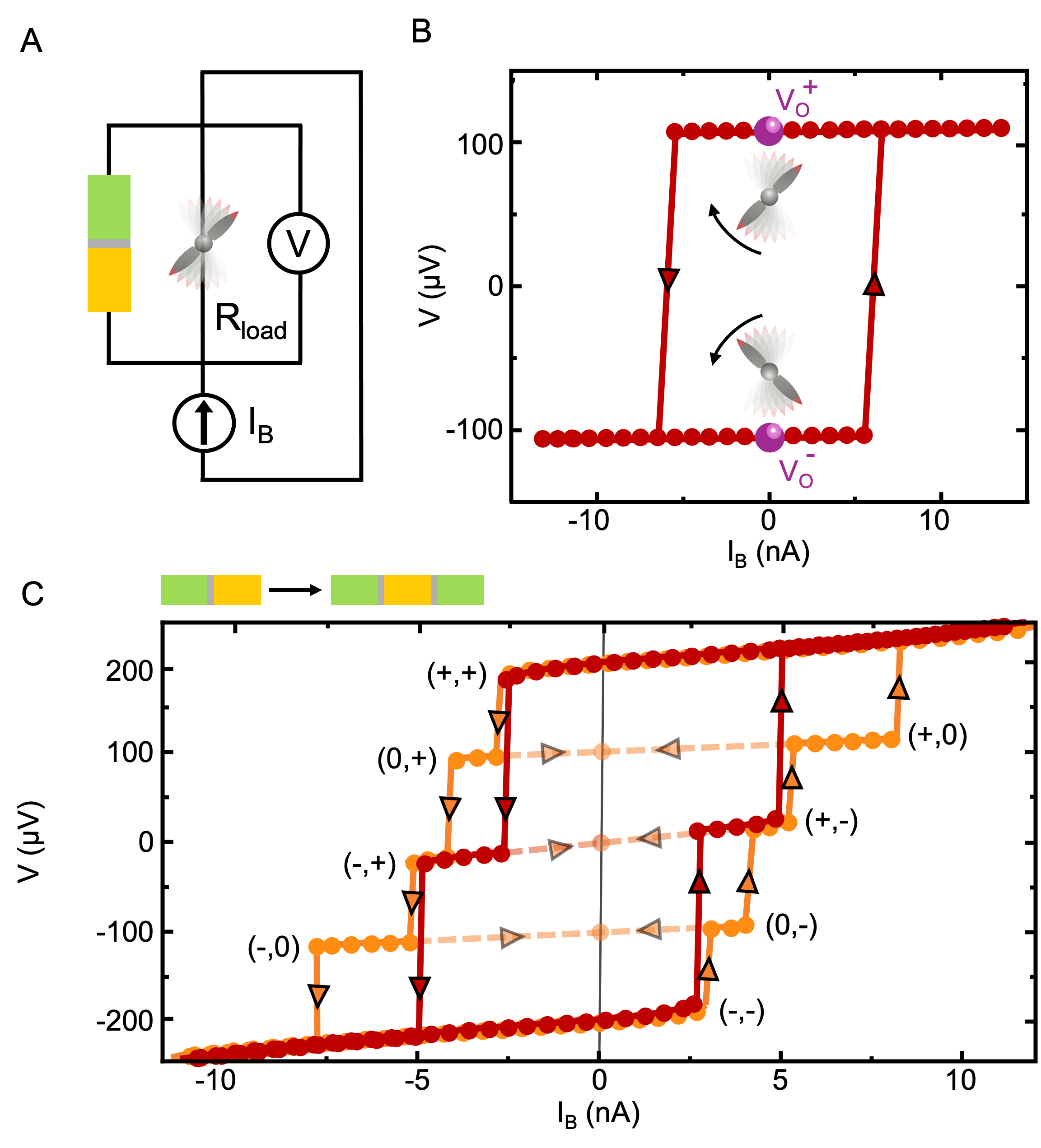}
\caption{\textbf{A heat-powered automaton under load.}
(A) Single junction loaded in parallel by $R_{\rm load}$ (drawn as a propeller) and biased by a current $I_B$, with the output voltage $V$ measured across it.
(B) Measured $V$ versus $I_B$ at $B^*$ for $R_{\rm load}=6~\mathrm{k}\Omega$ and $P_{\rm in}=3$~nW. Purple markers denote the output levels $V_{\rm O}^{\pm}$ at $I_B=0$, and propeller icons denote the output polarity.
(C) Output of the two-junction device of Fig.~\ref{fig:pair} for $R_{\rm load}=100~\mathrm{k}\Omega$ and $P_{\rm in}=0.7$~nW, at $B^*$ (red) and at $B^*-\delta B'$ with $\delta B'=1.6$~mT (orange). The icon marks the passage from the single junction to the pair. Transitions at negative (positive) $I_B$ occur during descending (ascending) sweeps. Labels give the signs of $V_L$ and $V_R$ in each state, and faded dashed lines show inner states carried back to $I_B=0$ by reversing the sweep.}
\label{fig:load}
\end{figure}

\section*{Heat-powered automaton}
A single junction loaded with a parallel resistor (Fig.~\ref{fig:load}A) already operates as a two-state machine~\cite{Germanese2022}. With $R_{\rm load}=6~\mathrm{k}\Omega$ and bias current $I_B$ at $B^*$ (Fig.~\ref{fig:load}B), its states lie where its characteristic meets the load line $I=I_B-V/R_{\rm load}$ (supplementary text S3). At $I_B=0$, the line crosses both generating branches where the characteristic is steep, giving two nearly flat output levels, $V_{\rm O}^{\pm}$. Each persists as $I_B$ changes sign, until its branch ends near the generation peak and the output jumps to the opposite polarity. The two levels thus form a hysteresis loop storing one bit, written by the sign of the last excursion and held at zero bias by the temperature difference. The junction meanwhile delivers about 1.9~pW to the load, roughly 80 times more than earlier devices~\cite{Germanese2022}.

Loaded with $R_{\rm load}=100~\mathrm{k}\Omega$ at $B^*$ (Fig.~\ref{fig:load}C, red), the pair of Fig.~\ref{fig:pair} keeps this bit and adds new ones. At large $|I_B|$, the only stable state is the concordant one with the sign of $I_B$, $(+,+)$ or $(-,-)$, where the two voltages add. Swept toward the opposite sign, it persists up to the outer generation peak, near $\pm 2V_P$ (Fig.~\ref{fig:pair}E), but there the output does not jump to the opposite polarity: the two junctions reverse one at a time. L reverses first, while R holds its voltage: the two nearly cancel, dropping the output close to zero in the discordant state $(-,+)$ or $(+,-)$. R reverses only at a larger $|I_B|$, bringing the pair to the opposite concordant state.
Detuning the field (orange) restores the single-junction peak near $V_P$, where one junction sits at zero voltage. The load-line solution also stops there, on both sides of the discordant state: $(0,+)$ and $(-,0)$ when descending, $(0,-)$ and $(+,0)$ when ascending.

A complete sweep traces the major hysteresis loop of Fig.~\ref{fig:load}C, which crosses $I_B=0$ only in the concordant states. Reversing the sweep within an inner state instead brings that state back to $I_B=0$, along a minor loop that rejoins the major one only at a further threshold (supplementary text S3). At $I_B=0$, each inner state leaves its own signature: a discordant state gives zero output but survives in the island potential, offset by $\pm V_S$ as in Fig.~\ref{fig:local}C, while a state with one junction at zero gives the output of its partner alone, as for the single junction of Fig.~\ref{fig:load}B. The state at rest thus records not only the sign of the last excursion but also which thresholds it crossed: four states at $B^*$, eight with the field detuned.

Each state corresponds to a distinct thermoelectric conversion configuration, specifying where in the circuit heat is converted into electrical power and which junctions deliver it to the load. The magnetic field determines the number of accessible states and shapes the switching thresholds through the critical currents, whereas junction asymmetry can make the two discordant states exit at distinct thresholds (supplementary text S3). Each state thus encodes the input history and exhibits a state-dependent response to subsequent stimuli, so that the coupled system operates as a heat-powered automaton~\cite{Liu2024,Kwakernaak2023}.

\section*{Conclusions}
We have scaled bipolar thermoelectric Josephson junctions to the picowatt regime and shown that two such junctions, coupled via a heated island, behave beyond a simple sum of their thermovoltages. The bias history sets the polarity of each junction and thereby selects which one acts as a power source, which acts as a dissipative element, and which carries a supercurrent. Thermal driving stabilizes multiple zero-bias configurations, and an applied magnetic field constrains the accessible states, so that the pair realizes a heat-powered automaton. These results indicate that multistability extends beyond voltage partitioning to the organization and allocation of thermoelectric energy conversion within the system.

Embedded in larger circuits, such junctions could implement this logic with greater functional complexity. For a linear array of $N$ junctions separated by alternating hot and cold islands, each junction can assume either polarity, yielding up to $2^N$ accessible configurations, or $3^N$ when the Josephson branch is also included. These thermally driven multistable superconducting circuits could thus provide a basis for persistent memory elements and sequential processing modules in superconducting electronics~\cite{Alam2023,Soloviev2017}, and, in the longer term, serve as scalable primitives for reservoir and neuromorphic computing~\cite{Liang2024,Schneider2018}.

The circuit elements also exhibit standalone functionality. A single junction delivers picowatts to a load, indicating that bipolar thermoelectricity may serve as a local power source for ultralow-power cryogenic circuitry~\cite{takeuchi2024microwave}. The concordant states, $(+,+)$ and $(-,-)$, further suggest junction arrays whose output scales with the number of junctions. Moreover, the floating island, thermally isolated by the tunnel barriers, could directly convert absorbed radiation into a thermoelectric readout~\cite{heikkila2018thermoelectric,paolucci2023highly}, with the hysteretic levels adding a field-tunable switching threshold once noise and relaxation are characterized~\cite{Pankratov2022}.

At smaller scales, where quantum effects matter, each junction, besides acting as its partner's load, also forms its electromagnetic environment, enabling photon-mediated energy exchange between the two~\cite{Hofheinz2011,Hofer2016}. Such photon exchange may itself drive the conversion. Indeed, a cold electromagnetic environment is predicted to induce bipolar thermoelectricity in a gap-asymmetric junction even with both electrodes at the same temperature~\cite{Antola2026QBT}. The operating principle considered here could thus extend to a purely radiative regime.

\section*{Acknowledgments}
We thank F. Taddei, M. Governale, L. Tosi and G. Trupiano for insightful scientific discussions.

\section*{Funding}
A.B. acknowledges support from the Italian Ministry of University and Research (MUR) under the PRIN 2026 ``Hybrid'' program, grant no.~2026HZ9RR3, ``Quantum Time: from foundation to applications'', from the project ``Thermoelectric effects in solid-state quantum devices based on multiterminal Josephson junctions'' of the CNR/CONICET (Italy/Argentina) bilateral agreement 2026--2027, from the CNR program STM2026, and from the CNR project QTHERMONANO.

\section*{Author contributions}
F.A. fabricated the devices, performed the experiments, analyzed the data, developed the circuit model, and wrote the initial draft.
M.C. contributed to the single-junction measurements and to the discussion of the results.
A.G. contributed to discussions on the design and fabrication of the devices and on the preliminary results.
A.B. contributed to the interpretation of the results and to the discussion of the circuit model.
G.D.S. provided support during the measurements and contributed to the discussion of the results.
F.G. conceived and supervised the project.
All authors revised and approved the final manuscript.

\section*{Competing interests}
The authors declare that they have no competing interests.

\section*{Data and materials availability}
All data needed to evaluate the conclusions in the paper are present in the paper or the supplementary materials. The data underlying the figures are deposited at Zenodo~\cite{dataZenodo}.

\bibliography{Bibliography}

% ============================================================
% Supplementary Material (single column)
% ============================================================
\clearpage
\onecolumngrid

\setcounter{section}{0}
\setcounter{figure}{0}
\setcounter{table}{0}
\setcounter{equation}{0}
\renewcommand{\thesection}{S\arabic{section}}
\renewcommand{\thefigure}{S\arabic{figure}}
\renewcommand{\thetable}{S\arabic{table}}
\renewcommand{\theequation}{S\arabic{equation}}
\renewcommand{\theHsection}{S\arabic{section}}
\renewcommand{\theHfigure}{S\arabic{figure}}
\renewcommand{\theHtable}{S\arabic{table}}
\renewcommand{\theHequation}{S\arabic{equation}}

\begin{center}
{\large\bfseries Supplementary Material\par}
\vspace{0.4em}
{\large\bfseries Multistable Thermoelectric Josephson Automata\par}
\vspace{0.6em}
Filippo Antola, Margherita Cervi, Angelo Greco, Alessandro Braggio, Giorgio De Simoni, and Francesco Giazotto
\end{center}
\vspace{1em}

\section*{Materials and Methods}
\label{sec:methods}

\subsection*{Device fabrication}
Devices were fabricated on Si substrates with 300~nm of thermally grown SiO$_2$, in two steps of maskless laser lithography (Raith picoMaster).

In the first step, a LOR~20B/S1805 bilayer resist stack was patterned to form a suspended mask. Tunnel junctions and integrated heaters were then defined in a single two-angle shadow-evaporation sequence in an electron-beam evaporator (base pressure $10^{-9}$~mbar, deposition rate 1~\AA/s). A 20~nm Al layer, forming the $\Delta_1$ electrode (the island in the two-junction devices), was deposited at $+48^\circ$ and oxidized in situ for 45~min in 12~mbar O$_2$ to form the AlO$_x$ tunnel barriers. Next, a Cu (11~nm)/Al (29~nm) bilayer was deposited at $-48^\circ$, with the Cu layer in direct contact with the barrier, forming both the $\Delta_2$ electrodes and the heaters, tunnel-coupled to the Al. A second in situ oxidation (10~min in 10~mbar O$_2$) then formed a clean oxide on the bilayer, which, after air exposure, acts as the barrier to the quasiparticle traps.

Following lift-off, a second lithography step using AZ1515 resist was aligned to the first pattern and developed in AZ Developer, which does not etch the Al films. The quasiparticle traps were then realized by sputter deposition of a 500~nm Cu film, followed by lift-off. The two-junction devices were fabricated together, with the same process, in a run separate from that of the single-junction device.

\subsection*{Measurement set-up}
All measurements were performed in a dry dilution refrigerator (Oxford Instruments Triton 200). The sample was thermally anchored to the mixing-chamber plate, and the bath temperature, measured with a RuO$_x$ thermometer, was maintained at 25~mK unless otherwise specified. Each electrical line was filtered by $\pi$-filters at room temperature and at the mixing chamber, the latter preceded by RC filters with $R=1~\mathrm{k\Omega}$ and $C=47~\mathrm{nF}$ for the single-junction device and $R=100~\Omega$ and $C=470~\mathrm{nF}$ for the two-junction devices.

Electrical transport was measured in a four-terminal configuration, with the voltage across the device measured by a voltage preamplifier (DL Instruments 1201) followed by a digital multimeter (Agilent 34401A). Current-biased characteristics were obtained by applying a voltage (Yokogawa GS200) across a series resistor much larger than the rest of the circuit, and voltage-biased characteristics by applying the voltage with the same source and recording the current with a current preamplifier (DL Instruments 1211). In the device with the voltage probe on the island, the local voltages $V_L$ and $V_R$ were measured between the island lead and each outer electrode with the same type of voltage preamplifier.

In each device, two heaters were connected in series to a single battery-powered floating source, so that the heating current entered the heated electrode through one heater and left through the other, with no net current through the junctions. The injected power was computed as $P_{\rm in}=I_hV_h$, where $V_h$ is the voltage drop across the two heaters. For the measurements of main-text Fig.~5, a load resistor was placed at room temperature in parallel with the device. The load $R_{\rm load}$ quoted throughout includes the series resistance of the filtered lines, and is therefore the load seen by the device.

\subsection*{Magnetic field and interference patterns}
An in-plane magnetic field was generated by a superconducting solenoid. The sample was mounted with its plane parallel to the solenoid axis, and the field was oriented perpendicular to the junction long axis $L$ (main-text Fig.~2A). The critical currents reported in the main text are switching currents, defined as the largest bias current reached over several current sweeps before a voltage above the noise floor appears. In the two-junction devices, $I_{cL}$ and $I_{cR}$ were extracted from the two successive switching events in the current--voltage characteristic (main-text Fig.~3B).

For a field parallel to the tunnel barrier and perpendicular to $L$, the flux threading the junction is $\Phi=B\,L\,d_{\rm eff}$, where $d_{\rm eff}$ is the effective magnetic thickness of the junction. Assuming a uniform current density in the rectangular junction, the critical current follows the Fraunhofer-like pattern~\cite{Rowell1963}
\begin{equation}
I_c(B) = I_{c0}\left|\frac{\sin(\pi \Phi/\Phi_0)}{\pi \Phi/\Phi_0}\right|,
\label{eq:fraunhofer}
\end{equation}
with period $\Delta B=\Phi_0/(L\,d_{\rm eff})$. The curve in main-text Fig.~2C is this expression with $I_{c0}=4.1~\mu$A, the measured maximum, and $\Delta B=6.9$~mT, which reproduces the width of the central lobe. A horizontal offset of 0.18~mT accounts for the residual field of the solenoid. For $L=15~\mu$m, the period gives $d_{\rm eff}\simeq20$~nm, of the order of the film thicknesses. It is the long side $L$ that brings the period down to a few millitesla. The measured pattern deviates slightly from the ideal one, with lower side lobes and outer minima at slightly lower fields. Both deviations indicate a current distribution along $L$ that is not perfectly uniform, as expected from a slightly non-rectangular junction shape or small inhomogeneities of the barrier.

The operating field $B^*$ was chosen as the interference minimum giving the strongest suppression of $I_c$. For the single-junction device, this is the second minimum, $B^*\simeq13.6$~mT. The two-junction devices, from a different fabrication run, show a larger period, $\Delta B\simeq10$~mT for the device of main-text Figs.~3 and~5 and $\Delta B\simeq9$~mT for that of main-text Fig.~4, and $B^*$ is the first minimum common to both junctions. These fields leave the superconducting gaps essentially unchanged, as the sum-gap threshold $V_\Sigma$ at $B^*$ differs from its zero-field value by less than a few percent.

\subsection*{Device parameters}
\begin{figure}
    \centering
    \includegraphics[width=\linewidth]{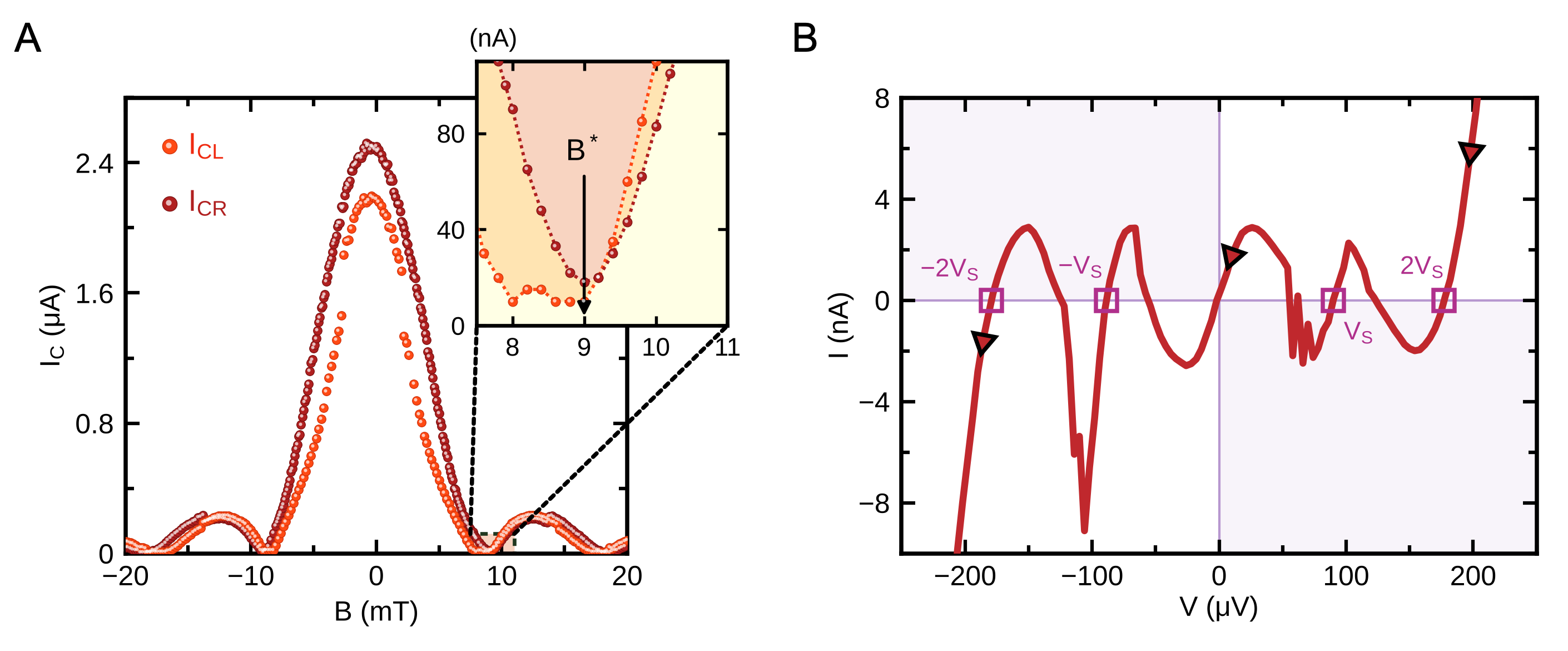}
    \caption{\textbf{Device with access to the island.} (A) Critical currents $I_{cL}$ and $I_{cR}$ versus $B$. Inset, minima near $B^*\simeq9$~mT, with shading as in main-text Fig.~3B. (B) Total current--voltage characteristic at $B^*$ under heating at $P_{\rm in}=1.9$~nW, swept from positive to negative bias (arrows). Purple shading highlights generation ($IV<0$), and squares mark the zero-current crossings at $\pm V_S$ and $\pm2V_S$.}
    \label{fig:probe}
\end{figure}

Here we summarize the main measured parameters of the single-junction device (main-text Fig.~2) and of the two two-junction devices, used in main-text Figs.~3 and~5 and in main-text Fig.~4, respectively. For the latter, which gives access to the island, we also show its interference pattern and a current--voltage characteristic under heating, both compatible with those of the other device (Fig.~\ref{fig:probe}).

The single-junction device consists of a $3.5\times15~\mu\mathrm{m}^2$ junction with a normal-state resistance $R_N\simeq37~\Omega$. Its Cu/Al bilayer has a critical temperature $T_c\simeq650$~mK, and at base temperature and zero field the critical current of the junction reaches $I_c\simeq4.1~\mu$A. The two tunnel heaters have a normal-state resistance of about 1.9~k$\Omega$ each and a critical current of about 24~nA.

The two-junction device of main-text Figs.~3 and~5 consists of two junctions of area $3\times15~\mu\mathrm{m}^2$ in series, sharing a floating Al island. Since the island is reached only through tunnel contacts, we report the total normal-state resistance of the pair, $R_{N,L}+R_{N,R}\simeq290~\Omega$. Its Cu/Al bilayer has a critical temperature $T_c\simeq700$~mK, and at base temperature and zero field the two junctions switch at different critical currents, $I_{cL}\simeq0.8~\mu$A and $I_{cR}\simeq1.1~\mu$A. The four tunnel heaters have normal-state resistances between 20 and 35~k$\Omega$ and a critical current of about 100~pA.

The two-junction device of main-text Fig.~4, fabricated in the same run and with the same bilayer critical temperature, differs from that of Figs.~3 and~5 by a superconducting lead contacting the island and by having two heaters instead of four. The lead gives access to each junction separately, with normal-state resistances $R_{N,L}\simeq57~\Omega$ and $R_{N,R}\simeq55~\Omega$. At base temperature and zero field the two junctions switch at $I_{cL}\simeq2.2~\mu$A and $I_{cR}\simeq2.5~\mu$A. The two tunnel heaters have a normal-state resistance of about 5.6~k$\Omega$ each and a critical current of about 4~nA.

Figure~\ref{fig:probe} characterizes this device at its operating point. The first interference minima of the two junctions are slightly misaligned, near 8.5~mT for L and 9.0~mT for R (Fig.~\ref{fig:probe}A). We take $B^*\simeq9$~mT, close to both minima, where $I_{cL}\simeq10$~nA and $I_{cR}\simeq18$~nA. The suppression is less effective than in the device of main-text Figs.~3 and~5, whose critical currents fall to a few nanoamperes at their common minimum (main-text Fig.~3C), and the residual critical currents account for the Josephson plateaus observed at $B^*$ in main-text Fig.~4B. Under heating at $B^*$, the total characteristic shows the features of main-text Fig.~3E (Fig.~\ref{fig:probe}B). It develops generation branches with $IV<0$ and crosses zero current near $\pm90$ and $\pm180~\mu$V (squares). These values are in a ratio of two, as expected for $\pm V_S$ and $\pm2V_S$, and $V_S\simeq90~\mu$V matches $V_R=-V_L$ at $V=0$ in main-text Fig.~4B. Additional features near $\pm55$ and $\pm125~\mu$V resemble those of the detuned characteristic of main-text Fig.~3E, consistent with the residual critical currents.

\clearpage
\section*{Supplementary Text}
\section{Quasiparticle thermometry and thermoelectric performance}
\label{note:thermometry}
This note describes how we extract the electronic temperatures of the hot and cold electrodes from the quasiparticle characteristics, and how we use them to obtain the nonlinear thermopowers and conversion efficiencies quoted in the main text.

\subsection*{Single-junction thermometry and conversion efficiency}
We fit the quasiparticle current--voltage characteristics of the single-junction device (main-text Fig.~2), measured at $B^*$, with a tunneling model. We label the heated Al electrode by 1 and the Cu/Al bilayer by 2, and describe each by a Fermi distribution at its electronic temperature $T_i$. The quasiparticle current is then
\begin{equation}
I_{\mathrm{qp}}(V)=\frac{1}{eR_N}\int_{-\infty}^{+\infty} n_1(E+eV,T_1)\,n_2(E,T_2)\left[f(E,T_2)-f(E+eV,T_1)\right]\,\mathrm{d}E,
\label{eq:qp_current}
\end{equation}
where $e>0$ is the elementary charge, $R_N$ the normal-state resistance and $f(E,T)=[1+\exp(E/k_{\mathrm B}T)]^{-1}$. We describe the normalized densities of states by the Dynes expression,
\begin{equation}
n_i(E,T_i)=\left|\operatorname{Re}\frac{E+\mathrm{i}\Gamma_i}{\sqrt{(E+\mathrm{i}\Gamma_i)^2-\Delta_i^2(T_i)}}\right|,
\label{eq:dynes_dos}
\end{equation}
where $\Delta_i(T_i)$ is the superconducting gap of electrode $i$ and $\Gamma_i$ a phenomenological broadening.

We fix $R_N=37\,\Omega$ to the normal-state resistance measured well above the sum-gap threshold, and calibrate the remaining parameters on the equilibrium characteristics. We assume the Al gap follows the BCS temperature dependence, with $T_{c1}=1.31\,\mathrm K$ and $\Delta_1(0)=1.764\,k_{\mathrm B}T_{c1}\simeq199\,\mu\mathrm{eV}$. The bilayer gap is obtained from the measured sum-gap threshold $V_\Sigma(T)$ as $\Delta_2(T)=eV_\Sigma(T)-\Delta_1(T)$, interpolated smoothly and constrained to vanish at the measured critical temperature of the bilayer, $T_{c2}=650\,\mathrm{mK}$, with $\Delta_2(0)\simeq71\,\mu\mathrm{eV}$. A global fit of the equilibrium characteristics, imposing $T_1=T_2=T_{\mathrm{bath}}$, gives $\Gamma_1\simeq0.02\,\mu\mathrm{eV}$ and $\Gamma_2\simeq8.7\,\mu\mathrm{eV}$, which we keep fixed under heating. The larger broadening of the bilayer phenomenologically captures part of the proximity-induced deviation of its density of states from the BCS form~\cite{hijano2023}.

\begin{figure}[t]
    \centering
    \includegraphics[width=\linewidth]{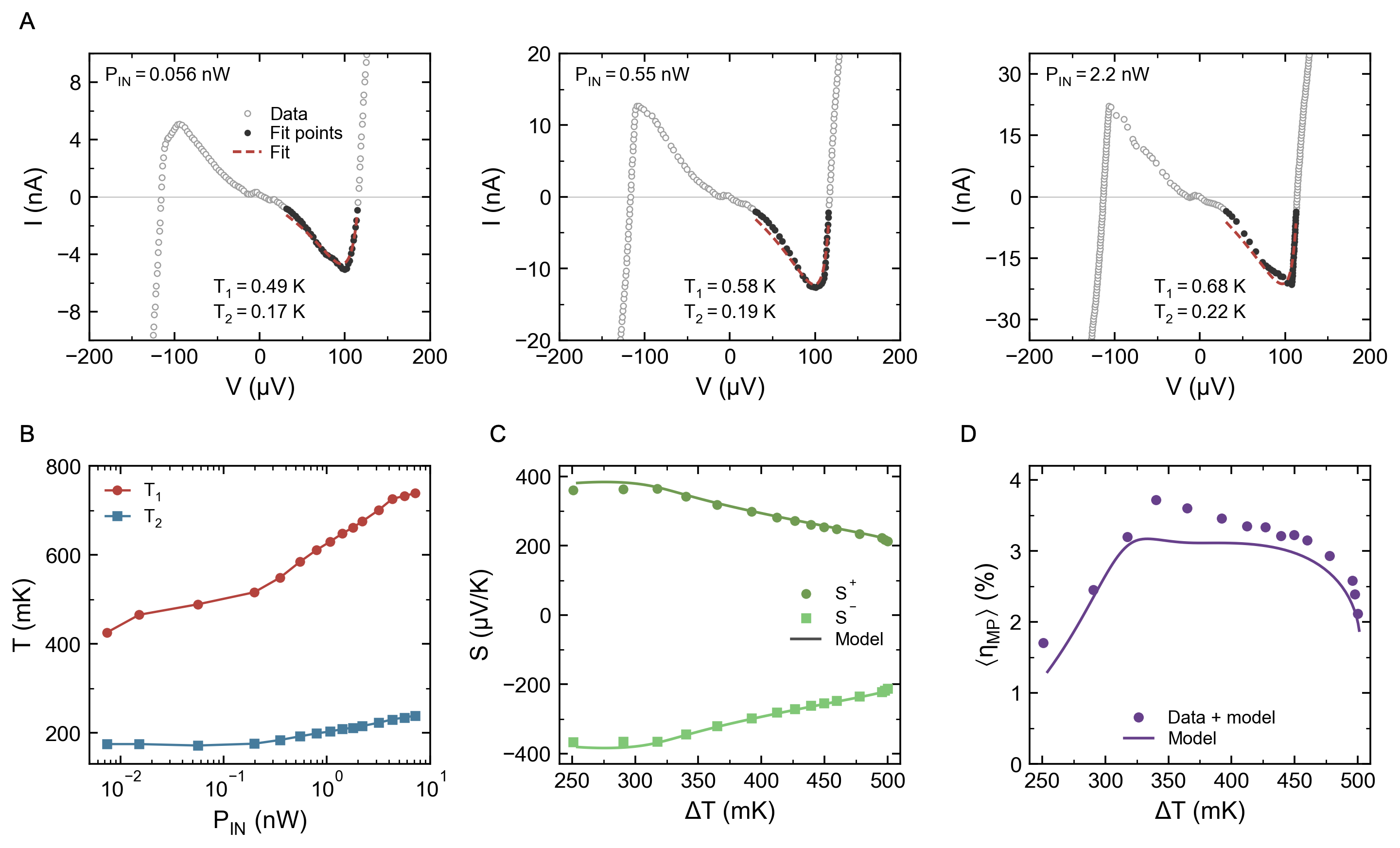}
  \caption{\textbf{Quasiparticle thermometry and thermoelectric performance of the single junction.}
(A) Current--voltage characteristics at the indicated heating powers $P_{\mathrm{in}}$: measured data (open gray circles), points used for thermometry (filled black circles) and fits (red dashed lines), with the extracted temperatures.
(B) Fitted temperatures $T_1$ and $T_2$ versus $P_{\mathrm{in}}$.
(C) Nonlinear thermopowers $S^\pm=V_S^\pm/\Delta T$ versus $\Delta T=T_1-T_2$. Symbols combine the measured thermovoltages with the fitted temperatures.
(D) Efficiency at maximum power averaged over the two polarities, $\langle\eta_{\mathrm{MP}}\rangle$, versus $\Delta T$. Symbols combine the measured maximum power with the heat current calculated at the same voltage.
In (C) and (D), lines are calculated from Eqs.~\eqref{eq:qp_current}--\eqref{eq:heat_current} with smooth interpolations of the fitted temperatures.}
\label{fig:sj_thermometry}
\end{figure}

Under heating, only the two electronic temperatures remain free (Fig.~\ref{fig:sj_thermometry}A,B). We determine them by least squares on the characteristic at positive voltage, corrected for the amplifier offsets, within the window $30\,\mu\mathrm V<V<V_S^+-1\,\mu\mathrm V$, where $V_S^+$ is the positive thermovoltage. The window excludes the residual Josephson contribution at low voltage and ends just before the zero-current crossing. Within it, we take $T_1$ and $T_2$ as independent of the bias.

The temperature difference $\Delta T=T_1-T_2$ gives the nonlinear thermopowers of the two polarities, $S^\pm=V_S^\pm/\Delta T$, which reach about $400\,\mu\mathrm{V\,K^{-1}}$ in magnitude (Fig.~\ref{fig:sj_thermometry}C), compared with up to $300\,\mu\mathrm{V\,K^{-1}}$ in the S/I/SN junctions of Ref.~\cite{Germanese2022}.

From the same densities of states and temperatures, we calculate the heat current leaving the hot electrode~\cite{Marchegiani2020PRL,Antola2026Review},
\begin{equation}
\dot Q_{\mathrm{hot}}(V)=\frac{1}{e^2R_N}\int_{-\infty}^{+\infty}(E+eV)\,n_1(E+eV,T_1)\,n_2(E,T_2)\left[f(E+eV,T_1)-f(E,T_2)\right]\,\mathrm{d}E,
\label{eq:heat_current}
\end{equation}
where $E$ is measured from the chemical potential of the cold electrode, and $\dot Q_{\mathrm{hot}}>0$ when heat leaves the hot electrode. For each polarity, we take the maximum generated power, $P^\pm_{\max}$, as the maximum of $-VI(V)$ on the measured characteristic, which in practice coincides with the generation peak at $V_P$ of main-text Fig.~2G, and evaluate $\dot Q_{\mathrm{hot}}$ at the voltage $V^\pm_{\mathrm{MP}}$ where it is reached. The efficiency at maximum power is then
\begin{equation}
\eta^\pm_{\mathrm{MP}}=\frac{P^\pm_{\max}}{\dot Q_{\mathrm{hot}}(V^\pm_{\mathrm{MP}})}.
\label{eq:efficiency}
\end{equation}

The efficiency averaged over the two polarities, $\langle\eta_{\mathrm{MP}}\rangle=(\eta_{\mathrm{MP}}^++\eta_{\mathrm{MP}}^-)/2$, reaches about $3.7\%$ (Fig.~\ref{fig:sj_thermometry}D, symbols), close to the efficiency obtained when the generated power is also calculated from Eq.~\eqref{eq:qp_current} (lines). The residual discrepancy likely reflects the proximity-modified density of states of the bilayer, which the Dynes form captures only in part~\cite{hijano2023}. These deviations shape the matching peak and hence the generated power.

\subsection*{Two-junction thermometry, heat balance and conversion efficiency}

\begin{figure}[t]
    \centering
    \includegraphics[width=\linewidth]{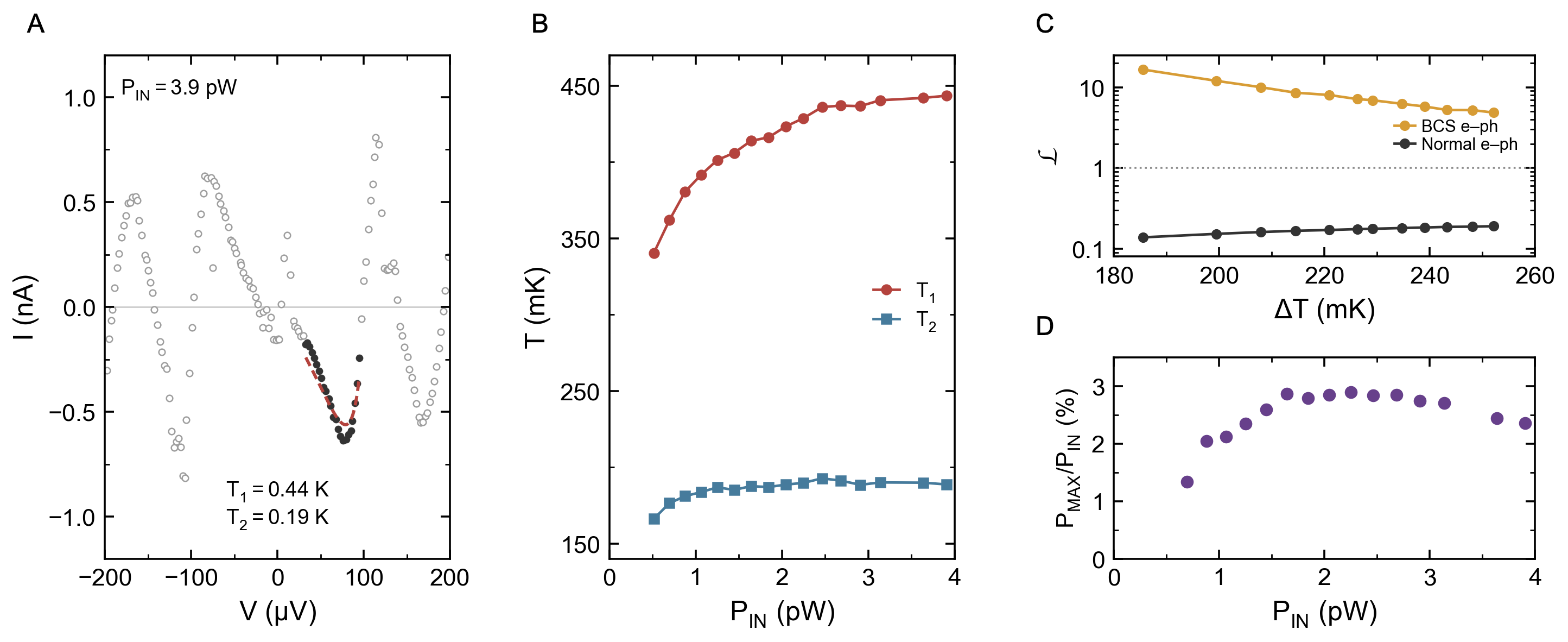}
\caption{\textbf{Thermometry, heat balance and conversion efficiency of the two-junction device.}
(A) Current--voltage characteristic at $P_{\mathrm{in}}=3.9\,\mathrm{pW}$, with symbols as in Fig.~\ref{fig:sj_thermometry}A and the extracted temperatures.
(B) Fitted temperatures $T_1$ and $T_2$ versus $P_{\mathrm{in}}$.
(C) Ratio $\mathcal{L}=\dot Q_{\mathrm{barriers}}/\dot Q_{e\text{-}ph}$ versus $\Delta T=T_1-T_2$, with the tunnel heat currents at the generation peak of the positive outer branch and $\dot Q_{e\text{-}ph}$ calculated for a BCS superconductor (orange) or with the normal-state law (black). The dotted line marks $\mathcal{L}=1$.
(D) Ratio of the maximum power generated on the positive outer branch, $P_{\max}$, to $P_{\mathrm{in}}$, both measured.}
\label{fig:dj_thermometry}
\end{figure}

For the two-junction device of main-text Fig.~3, we perform thermometry at $B^*$ and low heating powers, $P_{\mathrm{in}}\leq4\,\mathrm{pW}$. There, the characteristic shows a distinct inner branch, on which one junction develops the whole voltage while the other carries the current on its Josephson branch, as the current stays below its residual critical current. We describe this branch with the single-junction model above, calibrated on the equilibrium characteristics of this device, and fit it within the same window, $30\,\mu\mathrm V<V<V_1^+-1\,\mu\mathrm V$, where $V_1^+$ is the positive inner crossing (Fig.~\ref{fig:dj_thermometry}A). Assuming that the resulting temperatures of the island, $T_1$, and of the electrodes, $T_2$ (Fig.~\ref{fig:dj_thermometry}B), hold on the outer branch, $\Delta T=T_1-T_2$ gives both thermopowers of the main text, $S_1^\pm=V_1^\pm/\Delta T$ and $S_2^\pm=V_2^\pm/\Delta T$, with the crossing voltages read directly from the measured characteristics.

Generation requires the island to stay hotter than the electrodes, and only the heat that leaves through the junctions can contribute to the conversion, whereas the heat lost to phonons bypasses them. The floating island of the device of main-text Figs.~3 and~5 is therefore designed to be small, superconducting and contacted only through tunnel barriers, so that its coupling to the phonon bath is weak~\cite{Giazotto2006}. To check this, we compare the heat current leaving the island through the two junctions with its electron--phonon heat loss,
\begin{equation}
\mathcal{L}=\frac{\dot Q_{\mathrm{barriers}}}{\dot Q_{e\text{-}ph}}, \qquad \dot Q_{\mathrm{barriers}}=\dot Q_{\mathrm{hot},L}+\dot Q_{\mathrm{hot},R},
\label{eq:heat_partition}
\end{equation}
where each tunnel contribution is given by Eq.~\eqref{eq:heat_current}, evaluated at the generation peak of the positive outer branch. We calculate $\dot Q_{e\text{-}ph}$ with its expression for a BCS superconductor, including scattering and recombination processes~\cite{Timofeev2009}, assuming a uniform electronic temperature $T_1$ in the island, a phonon temperature $T_{\mathrm{ph}}=25\,\mathrm{mK}$, an Al volume $\mathcal{V}=7.125\,\mu\mathrm{m}^3$ and an electron--phonon coupling constant $\Sigma=0.2\,\mathrm{nW\,\mu m^{-3}\,K^{-5}}$~\cite{Giazotto2006}. Over the range considered, $\mathcal{L}\simeq5$--$17$ (Fig.~\ref{fig:dj_thermometry}C): the heat leaves the island mainly through the junctions.

To isolate the role of superconductivity, we replace only the electron--phonon heat current with its normal-state expression, $\dot Q_{e\text{-}ph}^{\mathrm N}=\Sigma\mathcal{V}\left(T_1^5-T_{\mathrm{ph}}^5\right)$, leaving temperatures and tunnel heat currents unchanged. $\mathcal{L}$ then falls below 1 (Fig.~\ref{fig:dj_thermometry}C, black): under the normal-state law the island would lose most of its heat to phonons, and it is superconductivity that keeps this channel weak.

Since most of the heat leaves the island through the junctions, we quantify the conversion of the two-junction device directly by the ratio $P_{\max}/P_{\mathrm{in}}$, where $P_{\max}$ is the maximum of $-IV$ on the positive outer branch. Unlike $\eta_{\mathrm{MP}}$ of the single junction, this ratio involves only measured quantities, and it reaches about $3\%$ at $P_{\mathrm{in}}\simeq2$--$3\,\mathrm{pW}$ (Fig.~\ref{fig:dj_thermometry}D). Part of $P_{\mathrm{in}}$ is lost to phonons, and part is dissipated in the heaters without reaching the island: the ratio is therefore a lower bound on $P_{\max}/\dot Q_{\mathrm{barriers}}$, the efficiency with respect to the heat that actually flows through the junctions.

\clearpage
\section{Bias partition and its memory}
\label{note:partition}
We model the two-junction devices (main-text Figs.~3 and~4) as a circuit to show why a fixed bias admits several partitions between the junctions, which are stable, and how the bias history selects among them. Each junction is described by its quasiparticle current, Eqs.~\eqref{eq:qp_current} and \eqref{eq:dynes_dos}, neglecting the Josephson current, strongly suppressed at $B^*$. The model aims at the qualitative behavior of the devices, with realistic parameters listed in the caption of Fig.~\ref{fig:asymmetric_states}.

\subsection*{Stationary partitions and their stability}
With the island electrically floating, the junctions carry the same current and their voltages add up to the bias,
\begin{equation}
I_L(V_L)=I_R(V_R),\qquad V_L+V_R=V.
\label{eq:partition}
\end{equation}
Fixing $V$ leaves the current free, so that each junction acts as the load of the other. Since a characteristic with absolute negative conductance is multivalued under current bias~\cite{Germanese2022}, Eq.~\eqref{eq:partition} can admit several partitions $(V_L,V_R)$ at the same $V$.

\begin{figure}[t]
    \centering
    \includegraphics[width=0.8\linewidth]{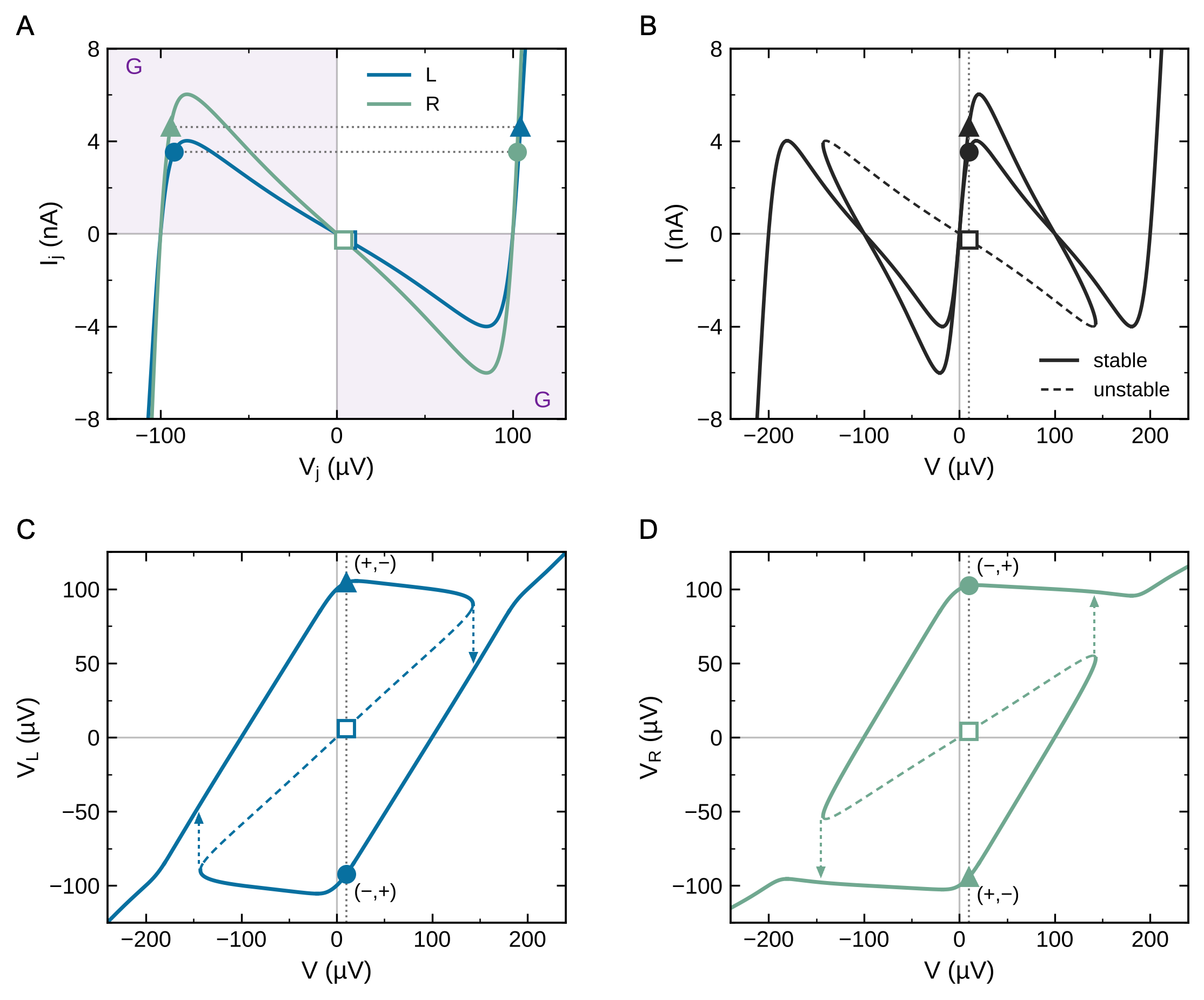}
    \caption{\textbf{Stationary voltage partitions of an asymmetric pair.} (A) Quasiparticle characteristics of the two junctions for $R_L/R_R=1.5$. Pairs of identical symbols mark the three solutions of Eq.~\eqref{eq:partition} at $V^*=10~\mu$V, joined by dotted lines at their common current; filled symbols denote stable solutions, open symbols the unstable one. Shading marks generation ($IV<0$). (B) Total current versus bias for all stationary partitions. Symbols mark the three solutions at $V^*$ (dotted line). (C and D) Corresponding junction voltages $V_L$ (C) and $V_R$ (D). Solid and dashed lines denote stable and unstable partitions throughout, and dotted arrows the jumps where a stable branch ends. All calculations use $R_L+R_R=290~\Omega$ and $T_c=0.70$~K for Cu/Al as measured, BCS gaps with $T_c=1.31$~K for Al and $\Delta(0)\simeq85~\mu$eV for Cu/Al, Dynes broadenings $\Gamma_{\mathrm{Al}}=1~\mu$eV and $\Gamma_{\mathrm{Cu/Al}}\simeq8.7~\mu$eV, and $T_{\mathrm{hot}}=700$~mK on the island and $T_{\mathrm{cold}}=250$~mK on the electrodes.}
    \label{fig:asymmetric_states}
\end{figure}

Figure~\ref{fig:asymmetric_states}A shows the junction characteristics for $R_L/R_R=1.5$, an asymmetry larger than in the experiment, chosen to separate the two curves. Each partition defines a branch of the total characteristic (Fig.~\ref{fig:asymmetric_states}B), so that the same bias can correspond to several branches. At $V^*=10~\mu$V, for example, Eq.~\eqref{eq:partition} has three solutions (circle, triangle and square), each a pair of points on the junction characteristics at a common current (Fig.~\ref{fig:asymmetric_states}A). In two of them, the junctions hold nearly opposite voltages, close to their thermovoltages, and, as in the main text, we label them by the signs of $(V_L,V_R)$, here $(-,+)$ and $(+,-)$ (Fig.~\ref{fig:asymmetric_states}C,D). In the third (square), the junctions share the bias almost equally.

Not all solutions are stable. If $V_R$, the island potential relative to the right electrode, deviates by $\delta V$ at fixed $V$, the junction currents become unbalanced and a net current $(G_L+G_R)\,\delta V$, with $G_i=\mathrm{d}I_i/\mathrm{d}V_i$, flows out of the island. For $G_L+G_R>0$, this current restores the partition, otherwise it amplifies the deviation. The criterion involves the sum of the junction conductances, not the differential conductance of the series, which can be negative for a stable partition, and coincides with that of a single junction loaded by a conductance $G_R$~\cite{Marchegiani2020PRL}. Following the partitions as a function of $V$ (Fig.~\ref{fig:asymmetric_states}C,D), the two discordant ones are both stable for $|V|\lesssim150~\mu$V, while the equal sharing (squares) is not. Near $V=0$ their currents nearly coincide (Fig.~\ref{fig:asymmetric_states}B): the terminal characteristic cannot distinguish them, and they differ only in the sign of each junction voltage, that is, in the island potential.

\subsection*{Selection by the bias history}
Which partition is occupied depends on how the bias is reached. Starting from large positive $V$, where $(+,+)$ is the only partition, the pair follows a stable branch as $V$ decreases (Fig.~\ref{fig:asymmetric_states}C,D). First $V_L$ crosses zero while $V_R$ stays near its thermovoltage, so that the pair reaches $V=0$ in $(-,+)$. Then $V_R$ crosses zero, bringing the pair to $(-,-)$, until near $-145~\mu$V the branch meets the unstable one and ends, and the pair jumps to the only remaining stable partition (dotted arrows). The reverse sweep mirrors this sequence and crosses $V=0$ in $(+,-)$. The sweep direction thus selects the state at zero bias, because in both directions the more resistive junction, L, swings first.

To check this picture under an actual sweep, we integrate the charge balance of the island,
\begin{equation}
(C_L+C_R)\frac{\mathrm{d}V_R}{\mathrm{d}t}=C_L\frac{\mathrm{d}V}{\mathrm{d}t}+I_L(V-V_R)-I_R(V_R),
\label{eq:island_dynamics}
\end{equation}
where the first term on the right is the displacement current driven by the bias and the others are the tunnel currents into and out of the island. We take $C_L=C_R=3$~pF, of the order expected for AlO$_x$ junctions of this area~\cite{Prunnila2010}. Further contributions, such as a capacitance to the substrate or to a gate, would change only the time scale, not the partitions or the jumps between them. The bias is swept from $+800$ to $-800~\mu$V and back at $1~\mu\mathrm{V}\,\mu\mathrm{s}^{-1}$, with $R_L/R_R=1.1$, which yields a total characteristic close to the measured one (main-text Fig.~3E).

\begin{figure}[t]
    \centering
    \includegraphics[width=0.9\linewidth]{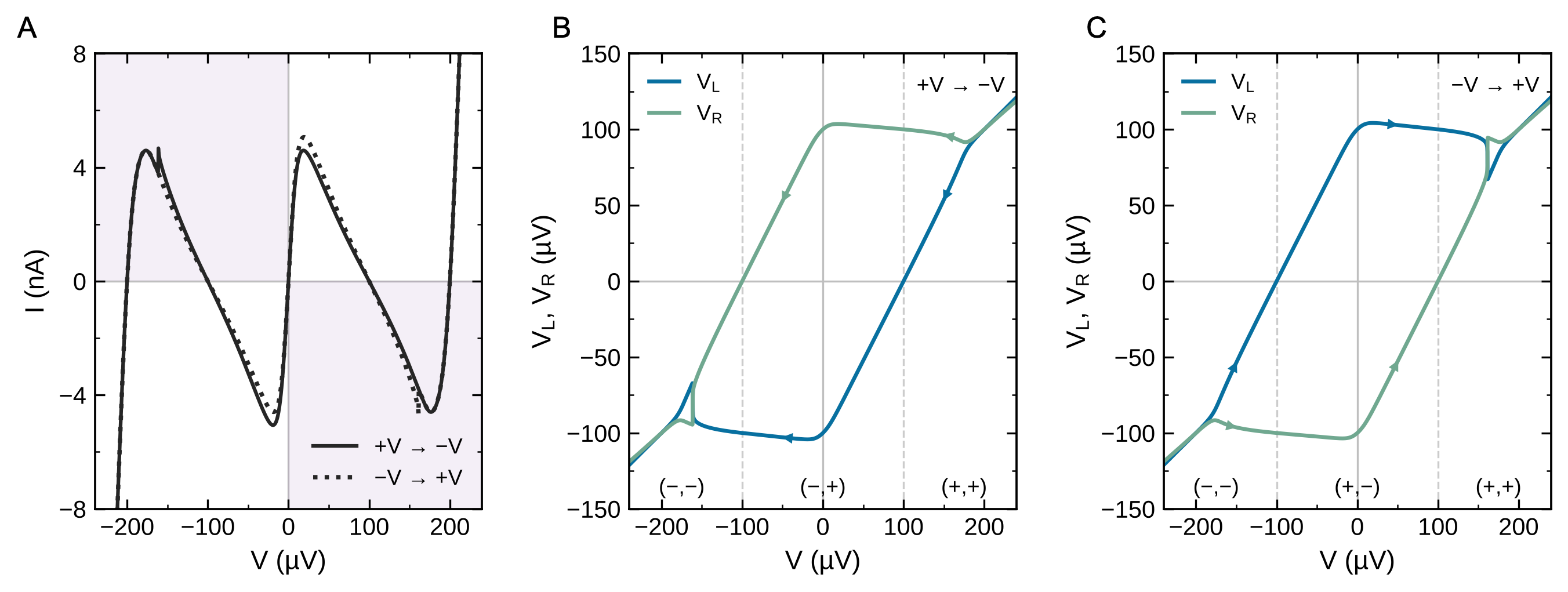}
    \caption{\textbf{Sweep-dependent voltage partition.}
    (A) Calculated total current versus bias for the descending (solid) and ascending (dotted) sweeps, for $R_L/R_R=1.1$. Shading marks generation ($IV<0$).
    (B and C) Junction voltages $V_L$ and $V_R$ along the descending (B) and ascending (C) sweeps. Arrows indicate the sweep direction and labels give the signs of $(V_L,V_R)$, with dashed lines near $\pm V_S$ separating the states. Other parameters as in Fig.~\ref{fig:asymmetric_states}.}
    \label{fig:voltage_sweeps}

    \vspace{1em}

    \includegraphics[width=0.9\linewidth]{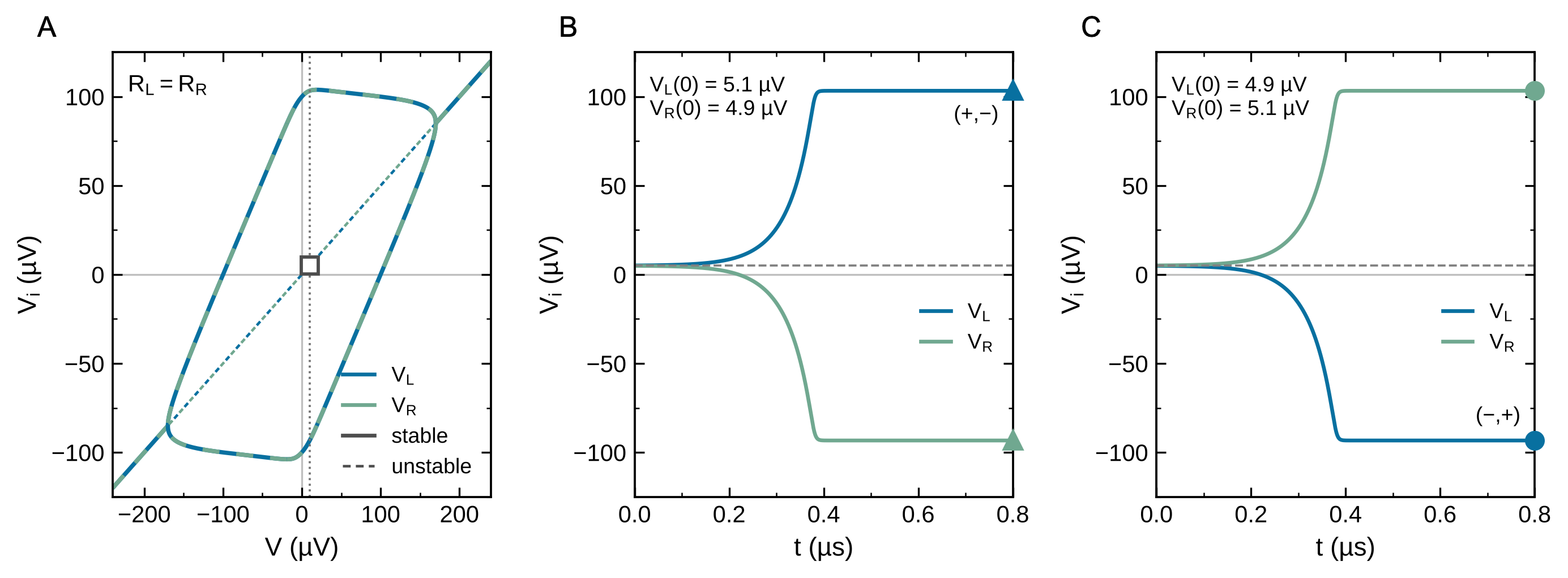}
    \caption{\textbf{Symmetry breaking for identical junctions.}
    (A) Stationary junction voltages versus bias for $R_L=R_R$. The solutions for $V_L$ and $V_R$ coincide by symmetry and are drawn in both colors. Solid and dashed lines denote stable and unstable partitions, and the square marks the equal sharing at $V=10~\mu$V (dotted line).
    (B and C) Evolution at this bias from $(V_L,V_R)=(5.1,4.9)~\mu$V (B) and $(4.9,5.1)~\mu$V (C). The dashed line marks the unstable equal sharing, $V/2$, and symbols mark the final states, as in Fig.~\ref{fig:asymmetric_states}. Other parameters as in Fig.~\ref{fig:asymmetric_states}.}
    \label{fig:identical_junctions}
\end{figure}

Figure~\ref{fig:voltage_sweeps} confirms the sequence inferred from the stationary partitions. The total characteristic (Fig.~\ref{fig:voltage_sweeps}A) crosses zero current at $V=0$, near $\pm V_S$ and near $\pm2V_S$, as the measured one, and is almost the same for the two sweep directions, apart from small spikes near $\pm160~\mu$V, where the partition jumps, similar to features of the measured characteristic (main-text Fig.~3E). The memory of the sweep is instead evident in the junction voltages (Fig.~\ref{fig:voltage_sweeps}B,C). On the descending sweep the pair passes through $(+,+)$, $(-,+)$ and $(-,-)$ and rests at $V=0$ with $V_L=-V_R\simeq-V_S$, while the ascending sweep visits the mirror states and reaches $V_L=-V_R\simeq+V_S$.

Junction asymmetry is not required for this behavior. For identical junctions, the partitions are symmetric under $V_L\leftrightarrow V_R$ (Fig.~\ref{fig:identical_junctions}A), and the equal sharing $V_L=V_R=V/2$ is unstable wherever the characteristic has negative differential conductance, since $G_L+G_R=2I'(V/2)<0$. Stable partitions with $V_L\neq V_R$ then occur in pairs, related by the exchange of the junctions. Figure~\ref{fig:identical_junctions}B,C shows the evolution at fixed $V=10~\mu$V from two slightly unbalanced initial conditions: the imbalance grows, as $G_L+G_R<0$, until the pair settles into one of the two discordant partitions, selected by the sign of the initial perturbation. Asymmetry therefore does not create the discordant states, but decides which junction swings first and makes the selection reproducible.

When the critical currents are not fully suppressed, as away from $B^*$, a junction whose voltage crosses zero can lock onto its Josephson branch and remain there as long as the current it carries stays below its critical current. The zero crossings of $V_L$ and $V_R$ then become the plateaus $(0,+)$ and $(-,0)$ observed in main-text Fig.~4B, without changing the order of the states. The extent of each plateau is set by the critical currents, which the model does not include.

\clearpage
\section{Operation under load}
In the loaded configuration of main-text Fig.~5, the device is connected in parallel with a load resistor $R_{\mathrm{load}}$ and driven by a current $I_B$. The current through the device is then $I=I_B-V/R_{\mathrm{load}}$, where $V$ is the output voltage. We calculate the junction characteristics with the quasiparticle model of sections S1 and S2, valid at $B^*$.

\subsection*{Single-junction hysteresis}
\begin{figure}[h]
    \centering
    \includegraphics[width=\linewidth]{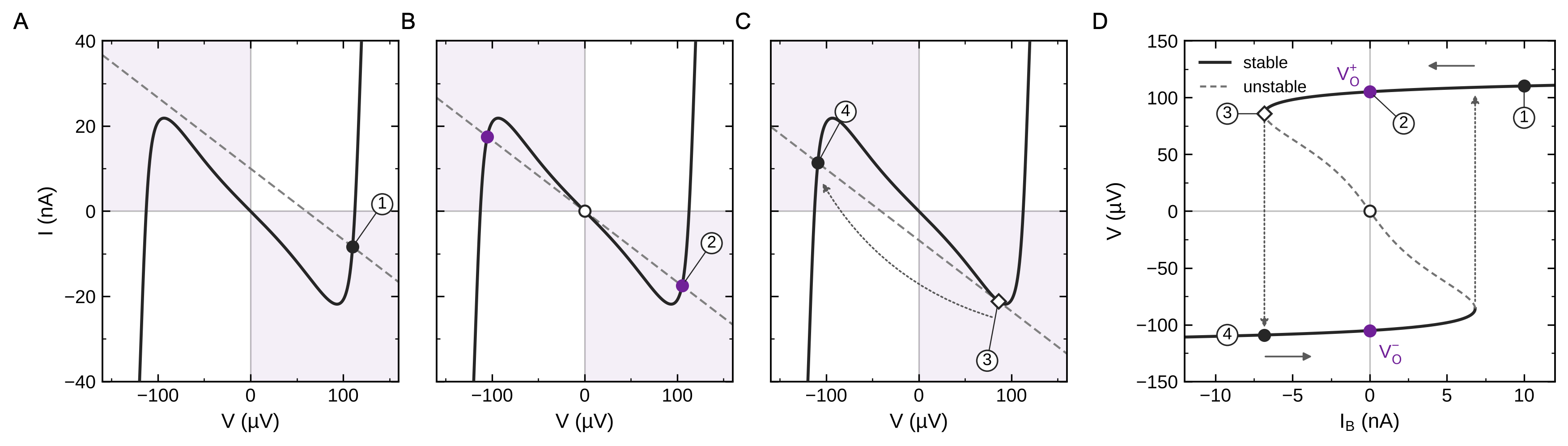}
    \caption{\textbf{Load-line construction for the single junction.}
(A to C) Junction characteristic (black) and load line $I=I_B-V/R_{\rm load}$ (dashed) for $R_{\rm load}=6~\mathrm{k}\Omega$, at $I_B=10$~nA (A), $0$ (B) and $-I_{\rm th}$ (C). Shading marks generation ($IV<0$). Filled and open circles denote stable and unstable intersections, and numbers follow a descending sweep. In (B), purple circles mark the output voltages $V_{\rm O}^{\pm}$ at $I_B=0$. In (C), the open diamond marks the tangency at $I_B=-I_{\rm th}$, where the positive state disappears, and the dotted arrow the jump to the negative branch.
(D) Output voltage versus $I_B$, with stable (solid) and unstable (dashed) branches, the points of (A) to (C), and arrows for the sweep directions and switching. Calculations use the fitted parameters of section S1, with $T_1\simeq700$~mK and $T_2\simeq223$~mK.}
\label{fig:single_junction_load}
\end{figure}

For a single junction, $I$ is set by its characteristic $I(V)$, so the operating points are the intersections of this characteristic with the load line $I=I_B-V/R_{\mathrm{load}}$, which $I_B$ shifts vertically without changing its slope (Fig.~\ref{fig:single_junction_load}A--C).

Figure~\ref{fig:single_junction_load} follows a sweep of $I_B$ from positive to negative values. At large positive bias, the load line crosses the characteristic only once, on the branch with $V>0$ (panel A, point 1), so the sweep selects the positive polarity. At $I_B=0$, the line passes through the origin and crosses both generating branches (panel B). The two outer intersections are stable, with opposite output voltages $V_{\rm O}^{\pm}$, while the origin is unstable. As for the pair under voltage bias (section S2), an intersection is stable when a small rise in $V$ makes the junction and the load draw more current than the source supplies, bringing $V$ back. This requires the sum of their conductances, $dI/dV+1/R_{\mathrm{load}}$, to be positive, whereas at the origin the characteristic falls more steeply than the load line. The junction reaches $V_{\rm O}^{+}$ continuously from point 1 (point 2), and this state remains stable. As $I_B$ becomes negative, the operating point moves toward the generation peak, until at $I_B=-I_{\rm th}$ the load line becomes tangent to the characteristic (panel C, point 3). There, $dI/dV=-1/R_{\rm load}$ and the stability condition fails: beyond this bias no state is left at positive voltage, and the junction jumps to the negative branch (point 4). The ascending sweep mirrors this sequence, closing the hysteresis loop of panel D, which reproduces the measured one of main-text Fig.~5B and the loop previously reported for a bipolar thermoelectric Josephson engine~\cite{Germanese2022}.

\subsection*{Major and minor loops of the pair}
\begin{figure}[t]
    \centering
    \includegraphics[width=\linewidth]{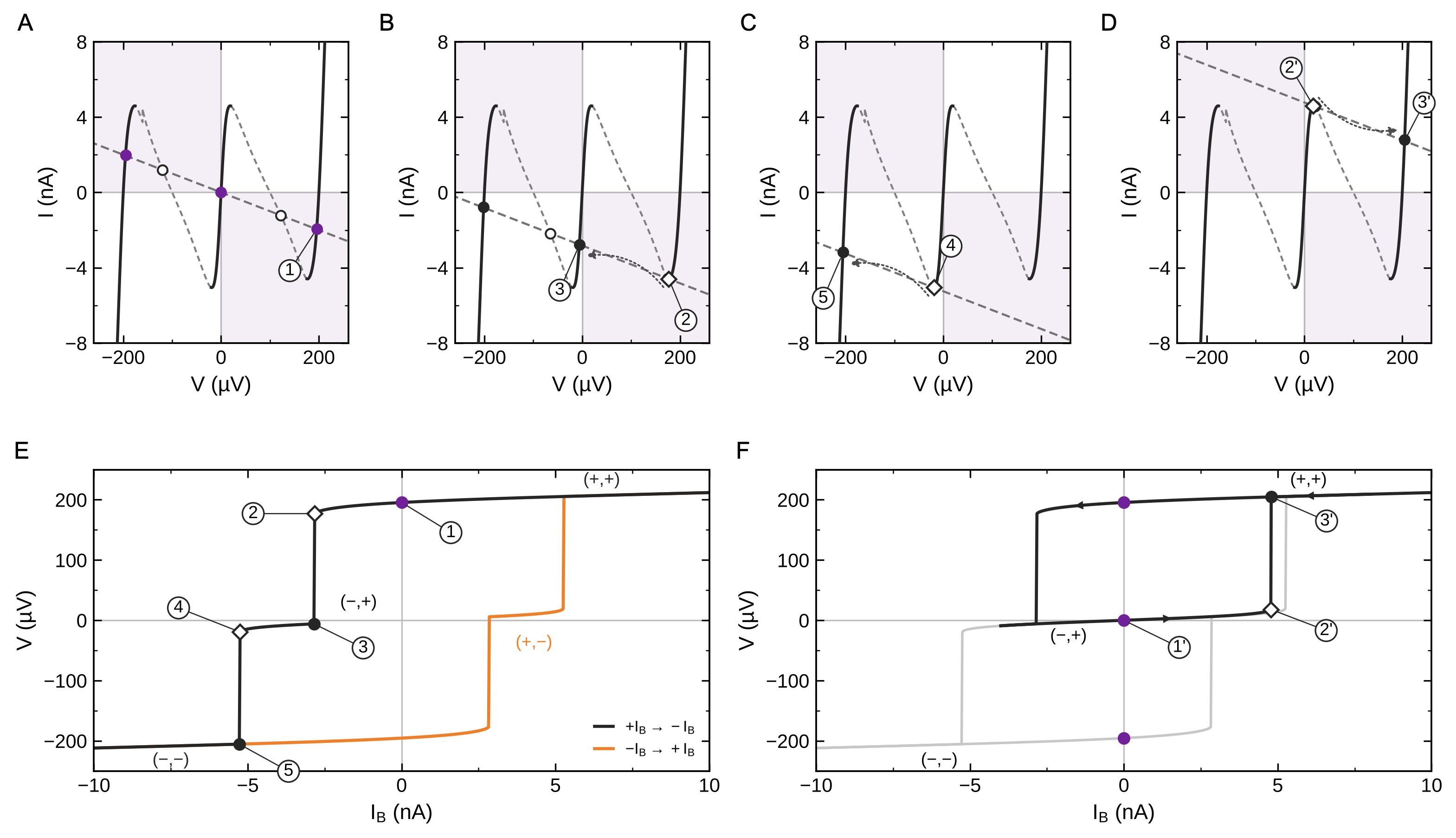}
    \caption{\textbf{Load-line construction for the two-junction device.}
(A to D) Device characteristic for a descending voltage sweep, with branches stable (solid) and unstable (dashed) under load, and load line for $R_{\rm load}=100~\mathrm{k}\Omega$ (gray dashed). The bias is $I_B=0$ (A), $-I_{\rm th,1}$ (B), $-I_{\rm th,2}$ (C) and the bias at which the minor loop rejoins the major one (D). Shading marks generation ($IV<0$). Filled and open circles denote stable and unstable intersections, and purple circles the stable states at $I_B=0$. Open diamonds mark where a branch ends, and dotted arrows the jumps.
(E) Major loop of the output voltage for descending (black) and ascending (orange) sweeps of $I_B$, with numbers following the descending sweep in (A) to (C).
(F) Minor loop (black), obtained by reversing the descending sweep within the discordant state, with the major loop in gray. Primed numbers mark the return sweep, also shown in (D).
Labels give the signs of $V_L$ and $V_R$. Calculations use the model and parameters of section S2, with $R_L/R_R=1.1$.}
\label{fig:double_junction_load}
\end{figure}

For the pair, the load line intersects the characteristic of the two junctions in series (Fig.~\ref{fig:double_junction_load}A), and, since $R_{\mathrm{load}}$ is much larger than the junction resistances, each junction must satisfy the stability condition of the single junction (solid lines). Unlike for the single junction, whose origin is unstable once the Josephson current is suppressed, the intersection at the origin is now stable at $I_B=0$: it corresponds to the two discordant states, $(-,+)$ and $(+,-)$, in which the junction voltages cancel.

Along a descending sweep from large positive bias, the pair stays in $(+,+)$ down to $I_B=0$ (Fig.~\ref{fig:double_junction_load}A, point 1). As $I_B$ becomes negative, the current through the pair approaches the generation peaks of the two junctions, which act as thresholds, as for the single junction. L, the less conductive junction, reaches its lower peak first, at $I_B=-I_{\rm th,1}$. There the $(+,+)$ branch ends and L alone reverses, bringing the pair to $(-,+)$, where the output nearly vanishes (panel B, points 2 and 3). R reverses only at $I_B=-I_{\rm th,2}$, when the current reaches its own peak, and the pair reaches $(-,-)$ (panel C, points 4 and 5). The output thus reverses in two steps (panel E). The ascending sweep follows the mirror sequence through $(+,-)$, and the resulting loop reproduces the one measured at $B^*$ in main-text Fig.~5C.

Once the pair is in a discordant state, reversing the sweep brings it back to $I_B=0$ (panel F, point 1'), where it rests with zero output and the island offset by $\pm V_S$, as measured in main-text Fig.~4B. Increasing $I_B$ further, the current reaches the generation peak of L, which reverses and brings the pair back to $(+,+)$ (panels D and F, points 2' and 3'). The sweep thus traces a minor loop, nested within the major one. Since the output is nearly zero in both discordant states, the current through the pair is close to $I_B$, and each discordant state ends at the generation peak of the junction that reverses: $(-,+)$ at the lower peak of L, $(+,-)$ at the higher peak of R. The two states thus exit at distinct thresholds (panel F), whose difference is set by the junction asymmetry and vanishes for identical junctions.

Away from $B^*$, the reopened critical currents give each junction a further stable branch at zero voltage, which it holds as long as the current stays below its critical current. Along a sweep, a junction reversing its polarity can then lock onto this branch, adding states such as $(0,+)$ and $(-,0)$, observed under load in main-text Fig.~5C. The critical currents, and hence the magnetic field, set the extent of these states and the current thresholds to leave them.

\end{document}